\documentclass[jasa]{article}

\RequirePackage{amsthm,amsmath,amsfonts,amssymb}
\RequirePackage[authoryear]{natbib}
\RequirePackage[colorlinks,citecolor=blue,urlcolor=blue]{hyperref}%
\RequirePackage{graphicx}%
\usepackage{mathrsfs}

\usepackage{tikz}
\usepackage{subfig}
\usepackage[hyphens]{xurl}
\usepackage{rotating}

\usepackage{hyperref}
\usepackage{amsmath, amsfonts, amssymb, mathrsfs}
\usepackage{dcolumn}
\usepackage{caption}

\usepackage{filemod}
\usepackage{natbib}
\usepackage[ruled, vlined]{algorithm2e}
\usepackage{floatrow}
\usepackage{setspace}
\usepackage{verbatim}

\usepackage{graphicx}

\title{A Bayesian Hierarchical Model Framework to Quantify Uncertainty of Tropical Cyclone Precipitation Forecasts}
\author{Stephen A. Walsh\thanks{Corresponding author: Department of Statistics, 
	Virginia Tech, {\tt walsh124@vt.edu}} \and Marco A.R. Ferreira \thanks{Department 
	of Statistics, Virginia Tech} 
	\and David Higdon\footnotemark[2]
	\and Stephanie Zick \thanks{Department 
	of Geography, Virginia Tech}}
	\date{July 19, 2022}

\begin{document}

\maketitle
\bigskip

\begin{abstract}
Tropical cyclones present a serious threat to many coastal communities around the world. Many numerical weather prediction models provide deterministic forecasts with limited measures of their forecast uncertainty. Standard postprocessing techniques may struggle with extreme events or use a 30-day training window that will not adequately characterize the uncertainty of a tropical cyclone forecast. We propose a novel approach that leverages information from past storm events, using a hierarchical model to quantify uncertainty in the spatial correlation parameters of the forecast errors  (modeled as Gaussian processes) for a numerical weather prediction model. This approach addresses a massive data problem by implementing a drastic dimension reduction through the assumption that the MLE and Hessian matrix represent all useful information from each tropical cyclone. From this, simulated forecast errors provide uncertainty quantification for future tropical cyclone forecasts. We apply this method to the North American Mesoscale model forecasts and use observations based on the Stage IV data product for 47 tropical cyclones between 2004 and 2017. For an incoming storm, our hierarchical framework combines the forecast from the North American Mesoscale model with the information from previous storms to create 95\% and 99\% prediction maps of rain. For six test storms from 2018 and 2019, these maps provide appropriate probabilistic coverage of observations. We show evidence from the log scoring rule that the proposed hierarchical framework performs best among competing methods.
\end{abstract}


\section{Introduction}
\noindent
Tropical cyclones (TCs) are some of the most costly and deadly natural disasters in the United States and across the world. The repercussions of these storms can take years to resolve, with TC rainfall being the primary culprit behind damage to inland communities \citep[e.g.,][]{cutter2014hurricane}. Recent studies suggest that intensification rates of TCs may be increasing \citep[e.g.,][]{bhatia2019}. Hurricanes are TCs that form in the Atlantic Ocean and achieve wind speeds of at least 33 meters per second; the Atlantic hurricane season of 2020 was the most active on record \citep{KlotzbachSummary}. To effectively prepare communities and allocate emergency services, understanding and quantifying uncertainty in model-based TC forecasts is paramount. In this paper, we develop a Bayesian hierarchical model framework to quantify uncertainty of TC precipitation forecasts.

Our novel hierarchical model combines the numerical weather prediction (NWP) TC forecasts with historical data to predict rainfall for TCs and to more realistically characterize the uncertainty in future predictions. We focus on precipitation that takes place 24 hours after an individual TC landfall and consider the error fields for each storm to be defined as the observational data minus the NWP forecast \citep{gel2004calibrated}. Our hierarchical model assumes that the error field for each TC follows a Gaussian process with a TC-specific set of parameters. We link the several TCs by assuming the parameters from each TC are realizations from a common linear model structure that may depend on known covariates. For example, in this paper we may assume that the prior distribution for the TC parameters depends on whether the storm makes landfall at the Atlantic, Florida, or Gulf of Mexico coastline.

We develop a Markov Chain Monte Carlo (MCMC) algorithm to explore the posterior distribution of the parameters of our hierarchical model. These include TC-specific parameters and hierarchical parameters related to the variability across different TCs. Here, we consider a training set of 47 TCs which made landfall between 2004 and 2017. Each TC’s set of data points ranges from 1,202 to 12,347, yielding a total of 308,013 data points for the 47 storms in the training set. In addition, analysis of spatial datasets with Gaussian processes is computationally expensive and usually scales cubically with sample size. Thus, a usual MCMC implementation of our hierarchical model would be infeasible. To deal with this massive data analysis problem, first we assume that the maximum likelihood estimate (MLE) of the parameters for each TC as well as the corresponding Hessian matrix contain all relevant information for that storm. We use these statistics to approximate the likelihood function for each TC. Finally, we develop an MCMC algorithm that uses these approximate likelihood functions to explore the posterior distribution of our proposed hierarchical model.

We use the output from the MCMC algorithm to obtain probabilistic forecasts for future storms. Specifically, we use the posterior sample of the hierarchical parameters to generate a sample from the prior distribution of the parameters of an incoming TC. With this sample, we use conditional sampling and the output of the NWP forecast to generate a sample of the precipitation fields. With this latter sample, we can compute precipitation probability maps to supplement the NWP forecast and better inform those affected by the incoming TC. We use storms from 2018 and 2019 as test data for this purpose. As we show in Section \ref{uq_future_storms}, our Bayesian hierarchical framework can also describe the plausible ranges of volume of water over crucial geographical regions (watersheds, floodplains, urban areas) that can adversely impact ecological systems (river flows and levels, flooding) and urban water systems.

NWP models use complex dynamical models based on atmospheric physics to describe future weather states. Specifically, here we consider NWP models that provide a single forecast without any uncertainty quantification. In particular, we do not consider probabilistic forecasting/data assimilation approaches (e.g. ensemble-based, variational) that might account for uncertainty in initial conditions, boundary conditions and parameterization choices to produce estimates of prediction uncertainty \citep{hamill2012noaa, bannister2017review}. We note that, similarly to what is discussed in \cite{gneiting2014probabilistic} and \cite{li2017review}, our methods may be adapted to statistically postprocess ensemble forecasts. For clarity of exposition we focus on NWP models that provide just one forecast.

Our Bayesian hierarchical framework may be considered a new statistical postprocessing approach. Statistical postprocessors are methods that seek to remove biases from both deterministic and ensemble forecasts. In addition to accounting for uncertainty in initial conditions and model uncertainty, statistical postprocessing offers a variety of methods to calibrate biased model output and illustrate uncertainty with predictive distributions \citep[see][for an extensive review]{li2017review}. 

When only a single NWP forecast is available, one method of upgrading to a probabilistic forecast can be achieved through the geostatistical output perturbation approach \citep{gel2004calibrated}. Their work uses pairs of previous forecasts and observations in a rolling window before the time of interest to learn about the discrepancies in the forecast and characterize the uncertainty by generating error fields with spatial covariance estimated by the data pairs. Other approaches that incorporate prior information in the form of climatological information include the Bayesian processor of forecast approach \citep{krzysztofowicz2008probabilistic} and work by \cite{schaake2007precipitation} and \cite{berrocal2008probabilistic}. Using climatology or a rolling window assumes that the commonplace weather patterns for that location will suffice to predict a future event. However, TCs are rare and extreme events and will likely not be characterized well by these approaches. Rather than using a rolling window or climatological information, our proposed framework uses a hierarchical model to account for forecast uncertainty for TCs.

Since the seminal work of \citet{gel2004calibrated}, a number of methodological extensions have been developed.  Perhaps the most prominent extensions account for multiple forecasting models; two of the most popular methods for Gaussian variables are Bayesian model averaging 
\citep{raftery2005using} and nonhomogeneous Gaussian regression \citep{gneiting2005calibrated}.

Additional extensions include accounting for non-Gaussian variables like precipitation \citep{sloughter2007probabilistic}, and the inclusion of spatial components \citep{berrocal2007combining, kleiber2011geostatistical, feldmann2015spatial}.

Recently, \cite{villarini2022probabilistic} proposed a method for generating probabilistic precipitation forecasts for TCs that made landfall in Louisiana. This method is applied to one particular region so the sample size of TCs (twelve) is relatively low. Our Bayesian framework provides spatially coherent uncertainty quantification (UQ) for rare and extreme weather events from a single NWP model output -- 
a use case of interest to many forecasting centers \citep[e.g.,][]{ko2020evaluation}. Additionally, our method can be applied generally for any TC in the contiguous United States (CONUS).

Many postprocessing approaches encounter some challenges when forecasting extreme events. \cite{bishop2008bayesian} found that the Bayesian model averaging approach has a problematic treatment of extreme weather and propose including climatological information to alleviate this issue. \cite{williams2014comparison} compare postprocessing methods for extreme events and find that the methods become less reliable as more extreme events are considered. Williams et al. note the Lorenz 96 model \citep{lorenz1996predictability} used in the study has short tails which will be easier to predict than real-world non-Gaussian variables and encourage development of postprocessing techniques specifically dedicated to extreme events. TC precipitation is some of the most extreme precipitation and will only exacerbate these challenges with extreme events. Another challenge related to many of these methods is the requirement of a rolling window (typically suggested to be around 30 days) of previous forecast/observation pairs to analyze characteristics of the model discrepancy. The rolling window is likely to contain errors for dry days and commonplace precipitation events that will not adequately characterize the errors corresponding to a TC. 

In contrast, by explicitly modeling these extreme events, our Bayesian hierarchical model framework circumvents these difficulties by quantifying the uncertainty within a NWP TC precipitation forecast based on past storm events. To illustrate our approach, we select the North American Mesoscale (NAM) model for our NWP model and create error fields for 47 TC events from 2004 to 2017. We produce an estimate of systematic biases found from the NAM and supplement future forecasts with uncertainty quantification in a novel approach that does not require reevaluating the weather model. 

Section 2 details the data preparation and processing procedure. We also introduce the hierarchical modeling framework and computations used to create our probability maps. Section 3 presents a simulation study which shows adequate estimation of spatial parameters used within the hierarchical model. Our modeling and computations are applied to the NAM and Stage IV data in Section 4. Section 5 details the UQ procedure for the test storms and compares our model to competing approaches, with the corresponding results from the logarithmic scoring rule. Section 6 provides discussion for future approaches and concluding remarks.

\section{Uncertainty Quantification}

\subsection{Precipitation Forecasts and Observations}
\label{dataproc_prep}

We study uncertainties of precipitation forecasts from the North American Mesoscale (NAM) model \citep{janjic2003nonhydrostatic, Rogers2009}, which is a NWP model run by the National Centers for Environmental Prediction (NCEP). NAM forecasts are available for download at \url{https://www.ncei.noaa.gov/products/weather-climate-models/north-american-mesoscale}. Forecasts from the NAM have a resolution of approximately 12km over a domain which covers CONUS. The NAM produces forecasts for numerous meteorological variables; we will work specifically with quantitative precipitation forecasts (QPFs) at times which correspond to TCs making landfall in CONUS. The NAM was selected for this study given its relatively long historical record; hurricane forecasts from the NAM date back to 2004 (when including the earliest forecasts under the original name of Meso-ETA model) and it is still operational at the time of writing.

To assess the level of uncertainty in the NAM forecasts, we use the NCEP Stage IV product \citep{Lin2005}, hereafter identified as Stage IV, as ground truth observations. Stage IV is available to download from \url{https://data.eol.ucar.edu/dataset/21.093}. Stage IV is a quality controlled quantitative precipitation estimate (QPE) that synthesizes information from radars and rain gauges across the United States. \cite{nelson2016assessment} mention that, although there is a general underestimation for Stage IV at higher rain rates, the biases and fractional standard errors both decrease  and correlation to rain gauges increases as the rain rate increases. Additionally, the eastern river forecast centers (which constitute our entire domain of interest) show the smallest fractional standard errors. For more comprehensive details of Stage IV and its many applications, see \cite{nelson2016assessment}. Out of three competing QPEs, \cite{villarini2011characterization} found that Stage IV was the most effective for accurately estimating TC precipitation. Multiple studies have evaluated different QPEs during TCs or heavy rainfall by comparing their performances to that of Stage IV \citep{jiang2008differences,habib2009evaluation, habib2009validation, zagrodnik2013investigation}. Studies by \cite{clark2010neighborhood} and \cite{yan2016evaluation} also evaluate NAM precipitation forecasts among other QPFs while using Stage IV as ground truth.

\begin{figure}%
    \centering
    {{\includegraphics[width=4.75in]{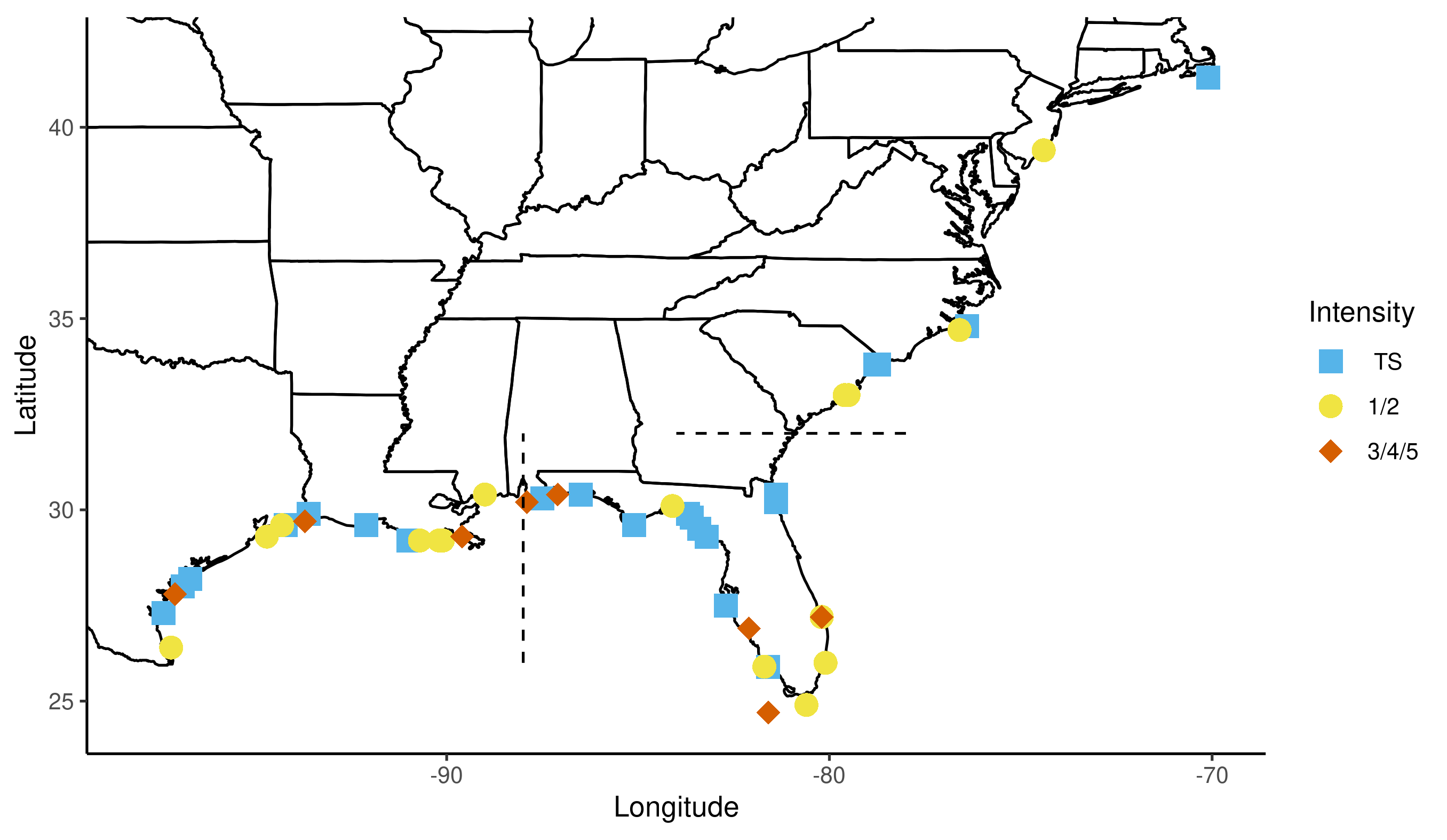} }}%
    \qquad
    \caption{Landfall locations and intensities of 47 TCs with tropical storm (TS) strength or greater from 2004 to 2017. Shapes indicate the Saffir-Simpson scale at the time of landfall. Dashed lines partition the Gulf, Florida, and Atlantic landfall regions as suggested by \cite{Jagger2006}.}%
    \label{fig:loc_int}%
\end{figure}

To study the most impactful portions of the storms, we focus on 24 hours of precipitation with NAM forecasts produced at either 0000 or 1200 Coordinated Universal Time (UTC). Specifically, if a storm makes CONUS landfall between 0600 UTC and 1800 UTC, we use the 1200 UTC forecast; otherwise, we use the 0000 UTC forecast. The same 24 hours of precipitation data are collected for Stage IV. The time and location of landfall for each of the storms are available in the second-generation hurricane database \citep[HURDAT2;][]{landsea2013atlantic} produced by the National Hurricane Center (NHC). \autoref{fig:loc_int} shows the $N=47$ TC landfall locations and intensities at time of landfall. 

To compare the NAM and Stage IV products, we interpolate both to a common coordinate reference system, the World Geodetic System 1984 (WGS84), with a spatial resolution of approximately 12km. To conserve the total volume of precipitation, we use a nearest neighbor interpolation scheme \citep{Accadia2003}. This interpolated grid is a rectangular array comprised of grid points both over CONUS and the neighboring oceans. Stage IV precipitation data collected over the ocean is inherently less reliable than data collected over land \citep{nelson2016assessment}. To account for this, we use a land-sea mask that excludes grid points over the ocean so that we only analyze precipitation over CONUS.

To focus solely on precipitation from the TC of interest, we employ two circular buffers with radii of 700km that remove all precipitation outside of the buffer. Commonly a 600km buffer is used (e.g., \citealt{Marchok2007,zick2016shape}); we choose to increase this to 700km to include precipitation for some of the larger or faster-moving storms. The two buffer centers are chosen to approximate the eye of the storm at 6 and 18 hours after landfall based on HURDAT2 information to adequately encompass the 24 hours of precipitation. We define the $i$th storm's buffer region $\mathcal{B}_i$ to be all grid points contained in either of the two overlapping circular buffers. Thus, all grid points in $\mathcal{B}_i$ will contain the $i$th storm's accumulated precipitation for the 24 hour time period; all other grid points are removed. We denote by $n_i$ the number of grid points in $\mathcal{B}_i$ which, as noted earlier, ranges from 1,202 to 12,347 for the 47 TCs in our training set. The union of all buffer regions establishes our common domain, $\mathcal{D}=\cup_{i=1}^{47}\mathcal{B}_i$, the set of all grid points over CONUS that are contained within at least one buffer region. For our training dataset, $\mathcal{D}$ has $n_\mathcal{D}=26,399$.

\begin{figure}
    \centering
    \tikzstyle{every node}=[font=\large]
\tikzstyle{process} = [rectangle, minimum width=2cm, minimum height=1cm, text centered, draw=black]
\tikzstyle{arrow} = [thick,->,>=stealth]

\noindent\resizebox{5.61893in}{!}{\begin{tikzpicture}[node distance=2cm]

\node (pro1) [process] {\begin{tabular}{c} Interpolate \\Stage IV \\ to the resolution \\ of NAM \end{tabular}};
\node (pro2) [process, right of=pro1, xshift=2cm] {\begin{tabular}{c} Mask \\ precipitation \\ over \\ocean \end{tabular}};
\node (pro3) [process, right of=pro2, xshift=2cm] {\begin{tabular}{c} Square root \\transformation of \\NAM and Stage IV \\precipitation data \end{tabular}};
\node (pro4) [process, right of=pro3, xshift=2cm] {\begin{tabular}{c}Employ two \\buffers of radius \\700km for 12-\\hour intervals \end{tabular}};
\node (pro5) [process, right of=pro4, xshift=2cm] {\begin{tabular}{c}Create error field: \\subtract NAM \\from the \\Stage IV \end{tabular}};

\draw [arrow] (pro1) -- (pro2);
\draw [arrow] (pro2) -- (pro3);
\draw [arrow] (pro3) -- (pro4);
\draw [arrow] (pro4) -- (pro5);

\end{tikzpicture}}
    \caption{Data preparation and processing steps.}
    \label{fig:data_proc}
\end{figure}
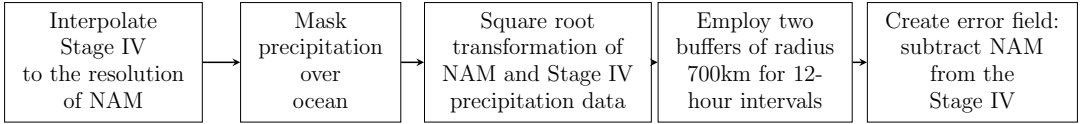

\begin{figure}[b]
    \centering
    \includegraphics[width=5.61893in]{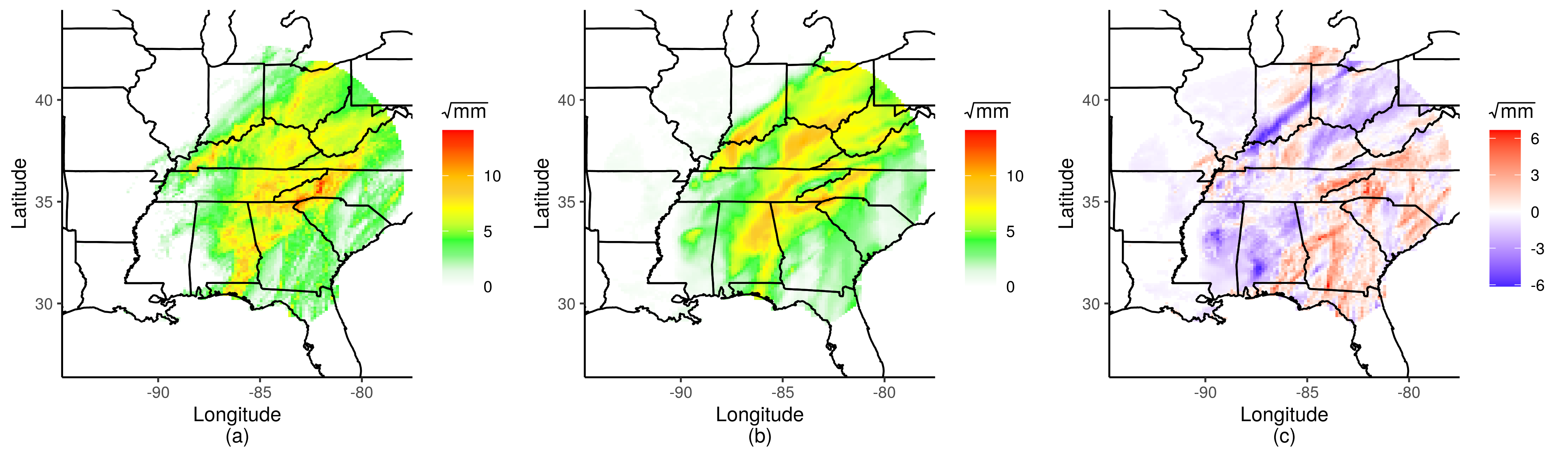}
    \caption{24 hour accumulated precipitation for Hurricane Nate, October 2017 (in mm) of (a) square root of Stage IV data, (b) square root of North American Mesoscale forecast, and (c) error field.}
    \label{fig:NAM_ST4_error}
\end{figure}

Precipitation information for both NAM and Stage IV are available in millimeters (mm). We originally used the log transformation to normalize and variance-stabilize the data, but this produced unrealistically high precipitation values when exponentiation was performed in the prediction phase. The square root transformation resolves this issue and is used instead. Let $\boldsymbol{m}_i$ and $\boldsymbol{o}_i$ be, respectively, the vector of square roots of NAM forecasts and the vector of square roots of Stage IV data for the $i$th storm within $\mathcal{B}_i$. We subtract the NAM forecast from the Stage IV observation to define the error field for each storm in our training dataset. Then, the vectorized error field for storm $i$ within $\mathcal{B}_i$ is defined as $\boldsymbol{y}_i=\boldsymbol{o}_i - \boldsymbol{m}_i$. The data processing procedure is summarized in \autoref{fig:data_proc}. We now have an error field for each of the 47 storms that we use to analyze biases and uncertainty within the NAM forecasts. Plots for $\boldsymbol{m}_i,\boldsymbol{o}_i,\boldsymbol{y}_i$ for $i=47$ are shown in \autoref{fig:NAM_ST4_error}, the data from Hurricane Nate in October 2017, the final storm in our training set. The plots for all storms are available in Figure S4 in the Supplementary Material \citep{walsh2022supplement}.

\subsection{Hierarchical Modeling for Error Fields}
\label{hiermodelforerrorfields}
Here we propose a hierarchical model for uncertainty quantification of the NAM forecasts. Specifically, we model the error field of the NAM forecast for each storm as a realization of a Gaussian process with storm-specific parameters. We then connect the different storms by assuming that the storm-specific parameters of the different storms are realizations from a common linear model structure.

We model $\boldsymbol{y}_i$ using a Gaussian process with a vector of covariance parameters $\boldsymbol{\lambda}_i=[\sigma^2_i,\phi_i]^T$ such that $\boldsymbol{y}_i \sim N(\boldsymbol{A}_i\boldsymbol{\mu},\boldsymbol{A}_i\boldsymbol\Sigma(\boldsymbol{\lambda}_i)\boldsymbol{A}_i^T)$. Based upon coverage probabilities for prediction storms shown in Section \ref{uq_future_storms}, we find the Gaussian assumption to provide reasonable results. The length of $\boldsymbol{y}_i$ is $n_i$, $\sigma^2_i$ is the marginal variance of the $i$th error field, and $\phi_i$ is the scale parameter establishing the empirical range of the $i$th error field (or the distance required for correlation to become negligible). We include $\boldsymbol{\mu}$, a vector of length $n_\mathcal{D}$, to model the mean process of these error fields, representing systematic biases that may be present within the TC forecasts. $\boldsymbol{A}_i$ is an incidence matrix (with dimension $n_i \times n_\mathcal{D})$ for the $i$th error field, indicating whether or not a particular grid point is within $\mathcal{B}_i$. That is, $\boldsymbol{A}_i$ is the mathematical equivalent to applying a buffer, reducing the domain of $\boldsymbol\mu$ from $\mathcal{D}$ to $\mathcal{B}_i$. 

Originally, we modeled $\boldsymbol{\Sigma}(\boldsymbol\lambda_i)$ with the more flexible Mat{\'e}rn covariance structure. We also included a nugget effect to describe microscale variation of the spatial process or measurement error. However, 21 of the 47 storms had an estimated nugget of 0 and only 7 of the storms had an estimated nugget above 0.1 (with a maximum nugget estimate of 0.25). For this reason we choose to continue with a more parsimonious model and set the nugget variance to 0. We also estimated the additional smoothness parameters for the Mat{\'e}rn model and found these estimates to have an average of 0.51 and all fell within [0.35, 0.74]. Given that the exponential covariance function is a special case of the Mat{\'e}rn with smoothness 0.5, we opt to use this more computationally efficient covariance structure. Thus, the covariance matrix for the error field $\boldsymbol{\Sigma}(\boldsymbol\lambda_i)$ is modeled by the exponential covariance function \citep{banerjee2014hierarchical}: 

\begin{equation}
    \boldsymbol\Sigma (\boldsymbol \lambda_i)_{j,k}= \textrm{Cov}(\boldsymbol{s}_j,\boldsymbol{s}_k)= 
      \sigma^2_i \exp\big(-||\boldsymbol{s}_j-\boldsymbol{s}_k||/\phi_i\big),
\end{equation}

\noindent where $||\boldsymbol{s}_j-\boldsymbol{s}_k||$ is the Euclidean distance between two grid points with locations $\boldsymbol{s}_j$ and $\boldsymbol{s}_k$. 

We wish to model the storms' spatial parameters with a Gaussian hierarchical model. Upon simulating 2,350 error fields with an exponential covariance function and a common true value for $\boldsymbol\lambda$, we see a correlation of 0.996 for the MLEs of $\sigma^2$ and $\phi$ as well as right-skewed distributions for each (see Figure S2 in the Supplementary Material, \cite{walsh2022supplement}). To reduce the correlation, we reparameterize our model and use $\boldsymbol\theta_i =[\theta_{i1},\theta_{i2}]^T= [\log(\sigma^2_i/\phi_i), \log(\sigma^2_i)]^T$ to model the covariance parameters. This reparameterization is motivated by results from  \cite{zhang2004inconsistent} and allows us to convert from a poorly behaved likelihood to one which is reasonably well approximated by a Gaussian distribution (see Figure S12 in the Supplementary Material, \cite{walsh2022supplement}). In our simulation study, this transformation makes the distributions approximately Gaussian and the correlation is -0.036 for $\theta_{i1}$ and $\theta_{i2}$; for details see Section \ref{simstudy} and Appendix C in the Supplementary Material \citep{walsh2022supplement}. Therefore, $\boldsymbol{\Sigma_\theta}$ will be nearly diagonal and our posterior distributions for $\boldsymbol\theta_i$ are well approximated by Gaussian distributions. 

Our resulting framework includes $\boldsymbol{\theta}_i\sim N(\boldsymbol{Bx}_i, \boldsymbol{\Sigma_\theta})$, where $\boldsymbol{B}$ is a matrix of regression coefficients, $\boldsymbol{x}_i$ is a vector of known regressors, and $\boldsymbol{\Sigma_\theta}$ is the corresponding covariance matrix for $\boldsymbol\theta_i$. One option for modeling the mean structure of $\boldsymbol\theta_i$ is to assume a common mean for each of the TCs. In this setting, $\boldsymbol{B}$ reduces to a vector of length two, with elements corresponding to the mean components of $\boldsymbol\theta_i$ and our known regressor reduces to a scalar with $x_i = 1$. In this setting we can define $\boldsymbol{\mu_\theta} \equiv \boldsymbol{B}x_i$ to represent a common mean across all TCs. $\boldsymbol{\Sigma_\theta}$ represents the covariance amongst components of each $\boldsymbol\theta_i$ across the storms. To avoid overfitting, we choose to estimate $\boldsymbol{\Sigma_\theta}$ as a common  covariance matrix across the different TC error fields.

Following work from \cite{Jagger2006}, we may categorize each of our storms by their landfall location, with three categories of Atlantic, Florida and Gulf storms delineated in \autoref{fig:loc_int}. This allows us to incorporate the influence of landfall location in the model. In the training dataset, we have 9 Atlantic, 21 Florida and 17 Gulf TCs. Here $\boldsymbol{x}_i$ is a vector of regressors indicating the landfall region; the first entry is a 1 which represents the Atlantic region as a baseline, with indicators in the second and third positions corresponding to effects from storms being in the Florida and Gulf regions, respectively. For this model $\boldsymbol{B}$ is a matrix of regression coefficients with the first column expressing the expected values of $\boldsymbol\theta_i$ if the $i$th storm were an Atlantic storm. The second and third columns represent the differences in these expectations if the storm were in the Florida or Gulf regions respectively. The rows of $\boldsymbol{B}$ correspond to the spatial parameters contained in $\boldsymbol\theta_i$. 

We implement the following general hierarchical model for the TC error fields:
\begin{equation}
    \boldsymbol{y}_i|\boldsymbol{\mu},\boldsymbol{\theta}_i \sim N(\boldsymbol{A}_i \boldsymbol{\mu},\boldsymbol{A}_i \boldsymbol{\Sigma}(\boldsymbol{\theta}_i)\boldsymbol{A}_i^T),
    \label{eqn:y_i}
\end{equation}
\begin{equation}
    \boldsymbol{\mu}|\boldsymbol{m, C} \sim N(\boldsymbol{m},\boldsymbol{C}),
    \label{eqn:condl_mu}
\end{equation}
\begin{equation}
    \boldsymbol{\theta}_i|\boldsymbol{B, \Sigma_\theta} \sim N(\boldsymbol{Bx}_i, \boldsymbol{\Sigma_\theta}),
    \label{eqn:theta_i}
\end{equation}
\begin{equation}
    \pi(\boldsymbol{B}) \propto 1,
    \label{eqn:priorB}
\end{equation}
\begin{equation}
    \boldsymbol{\Sigma_\theta} \sim IW(\nu_0, \boldsymbol{S}_0).
    \label{eqn:priorSigma_theta}
\end{equation}

\noindent Including Equation \eqref{eqn:condl_mu} allows us to model variability of the mean process $\boldsymbol{\mu}$. After employing two different prior specifications for $\boldsymbol\mu$, we found that setting $\boldsymbol\mu=0$ provided the best scores based on the logarithmic scoring rule \citep{gneiting2007strictly}. Therefore, for this application we specify $\boldsymbol\mu=0$. For more details on this, see Section \ref{uq_future_storms} and Appendix B in the Supplementary Material \citep{walsh2022supplement}. Equation \eqref{eqn:theta_i} is the distribution of the spatial parameters for each error field conditional on the prior distributions \eqref{eqn:priorB} and \eqref{eqn:priorSigma_theta}.  Both priors for $\boldsymbol{B}$ and $\boldsymbol{\Sigma_\theta}$ constitute vague hyperprior specifications, with a flat prior for $\boldsymbol{B}$ and a conjugate inverse Wishart prior for $\boldsymbol{\Sigma_\theta}$, which permits the use of an efficient Gibbs sampler. For hyperparameters of $\boldsymbol{\Sigma_\theta}$, we set $\nu_0=p+1$ where $p$ represents the dimension of $\boldsymbol{\theta}_i$; this contains reasonably vague prior information while ensuring the prior distribution is proper. We explore options for prior settings of $\boldsymbol{S}_0$ through sensitivity analysis and find that the results are rather sensitive to the choice of $\boldsymbol S_0$. Therefore we implement an empirical Bayes method of setting $\boldsymbol S_0 = \nu_0 \textrm{Cov}(\hat{\boldsymbol\Theta})$, where the $i$th row of $\hat{\boldsymbol\Theta}$ is $\hat{\boldsymbol\theta}_i^T$, the maximum likelihood estimate (MLE) of $\boldsymbol\theta_i^T$. This prior specification has been shown to outperform other common prior specifications for $\boldsymbol S_0$ when the variances of the parameters (e.g. $\boldsymbol\theta$) are small \citep{schuurman2016comparison}. From this hierarchical model, we can learn the posterior distributions for $\boldsymbol{B}, \boldsymbol{\Sigma_\theta}$, and  $\boldsymbol{\theta}_i, i \in \{1,\dots,N\}$.

\subsection{Computations}
\label{computations}

We perform computations in two steps. In the first step, we compute the MLEs of $\boldsymbol\theta_1, \dots, \boldsymbol\theta_N$ as well as the corresponding Hessian matrices. After that, we use the Gaussian approximation to the distribution of $\hat{\boldsymbol\theta}_1,\dots,\hat{\boldsymbol\theta}_N$ to construct approximate likelihood functions for each storm. In the second step, we use these approximate likelihoods combined with the hyperpriors \eqref{eqn:theta_i}, \eqref{eqn:priorB}, and \eqref{eqn:priorSigma_theta}
to build a Gibbs sampler for $\boldsymbol\theta_1,\dots,\boldsymbol\theta_N$, $\boldsymbol{\mu_\theta}$, and $\boldsymbol{\Sigma_\theta}$. As we explain in Section \ref{uq_future_storms}, we use the posterior sample of $\boldsymbol{\mu_\theta}$ and $\boldsymbol{\Sigma_\theta}$ and the hierarchical model specification to obtain probabilistic forecasts for future storms. 

For each error field $\boldsymbol{y}_i$, we estimate $\hat{\boldsymbol{\theta}}_i=[\hat{\theta}_{i1},\hat{\theta}_{i2}]^T$ using the maximum of the profile likelihood \citep[e.g.,][Chapter 5]{diggleribeiro2007}. Let $\boldsymbol{H}_i$ denote the negative of the Hessian from the full likelihood function for the $i$th storm. We derive an approximate asymptotic covariance matrix with $\boldsymbol{H}_i^{-1}$, which can be calculated analytically (see Appendix A in the Supplementary Material, \cite{walsh2022supplement}) or with numerical approximations derived from the \texttt{pracma} package in \texttt{R} \citep{pracma}. The Hessians from the two methods match to the third decimal place and the numerical approximation was found to be 2.5-10 times faster. Therefore we implement the numerical approximation within our framework. From the simulation study results in the Supplementary Material \citep{walsh2022supplement}, we see empirical evidence that the distributions of the MLEs of $\theta_1$ and $\theta_2$ are very well approximated by a normal distribution. In this application, we obtain good coverages for the prediction storms' precipitation, so these approximations are reasonable. 

We obtain posterior draws for each of the parameters of interest by using a Gibbs sampler. The dimensionality of the $i$th storm, $n_i$, carries the computational burden of  $\boldsymbol{A}_i\boldsymbol{\Sigma}(\boldsymbol\theta_i)\boldsymbol{A}_i^T$ within the Gibbs sampler. To address this, we assume that the MLEs $\hat{\boldsymbol\theta}_i$ and their corresponding asymptotic covariance matrices $\boldsymbol{H}_i^{-1}$ contain all useful information for the $i$th storm's data. This achieves a massive data reduction, where the dimension of the problem is reduced from 308,013 to just $N \times p=94$. Therefore, we have that $\mathcal{L}(\boldsymbol{y}_i|\boldsymbol{\theta}_i)=\mathcal{L}(\hat{\boldsymbol{\theta}}_i|\boldsymbol\theta_i, \boldsymbol{H}_i) \sim N(\boldsymbol{\theta}_i, \boldsymbol{H}_i^{-1})$ which enables the Gibbs sampler to complete 10,000 iterations after burn-in in 150 seconds. The joint posterior is shown below, where $\boldsymbol y_i$ is a zero-mean Gaussian process. 

\begin{align*}
    \pi(\boldsymbol{\theta}_i, \boldsymbol{B}, \boldsymbol{\Sigma_\theta}|\boldsymbol{y}_i, \boldsymbol{x_i})
    &=\pi(\boldsymbol{\theta}_i, \boldsymbol{B}, \boldsymbol{\Sigma_\theta}|\boldsymbol{\hat\theta}_i, \boldsymbol{x_i}) \\
    &= \mathcal{L}(\boldsymbol{\hat\theta}_i|\boldsymbol{\theta}_i,\boldsymbol{H}_i^{-1})
    \pi(\boldsymbol{\theta}_i|\boldsymbol{\boldsymbol{Bx_i},\Sigma_\theta})
    \pi(\boldsymbol{B})\pi(\boldsymbol{\Sigma_\theta})\\
    &\propto \bigg\{\prod_{i=1}^{N} \exp\Big(-\frac{1}{2}(\boldsymbol{\hat{\theta}_i}-\boldsymbol{\theta_i})^T\boldsymbol{H_i}(\boldsymbol{\hat{\theta}_i}-\boldsymbol{\theta_i})\Big)\bigg\}\\
    &\qquad \times\bigg\{\prod_{i=1}^{N} |\boldsymbol{\Sigma_\theta}|^{-1/2}\exp\Big(-\frac{1}{2}(\boldsymbol{\theta_i}-\boldsymbol{Bx_i})^T \boldsymbol{\Sigma_\theta}^{-1}(\boldsymbol{\theta_i-Bx_i})\Big)\bigg\}\\
    &\qquad \times1\times|\boldsymbol{\Sigma_\theta}|^{-(\nu_0+p+1)/2}\exp(-\text{tr}(\boldsymbol{S_0\Sigma_\theta^{-1}})/2)
\end{align*}
  
The full conditional distributions for the Gibbs sampler are shown below; derivations are in Appendix A in the Supplementary Material \citep{walsh2022supplement}. The dash ($-$) indicates all inputs of the joint posterior with the exception of the particular variable for which the full conditional is defined. These distributions are multivariate normal, inverse Wishart and matrix-variate normal, respectively.
\begin{align*}
    \pi(\boldsymbol\theta_i|-) &\equiv N\Big((\boldsymbol{H_i}+\boldsymbol{\Sigma_\theta}^{-1})^{-1}(\boldsymbol{H_i\hat{\theta}_i}+\boldsymbol{\Sigma_\theta}^{-1}\boldsymbol{Bx}_i),(\boldsymbol{H_i}+\boldsymbol{\Sigma_\theta}^{-1})^{-1}\Big)\\
    \pi(\boldsymbol{\Sigma_\theta}|-)&\equiv IW\Big(N+\nu_0,\sum_{i=1}^N (\boldsymbol\theta_i-\boldsymbol{Bx}_i)(\boldsymbol\theta_i-\boldsymbol{Bx}_i)^T+\boldsymbol{S}_0\Big)\\
    \pi(\boldsymbol{B}|-) &\equiv MN\Big((\sum_{i=1}^N\boldsymbol{\theta}_i\boldsymbol{x}_i^T)(\sum_{i=1}^N \boldsymbol{x}_i\boldsymbol{x}_i^T)^{-1}, \boldsymbol{\Sigma_\theta}, (\sum_{i=1}^N\boldsymbol{x}_i\boldsymbol{x}_i^T)^{-1}\Big)
\end{align*}

This method is feasible in an online forecasting context; the Gibbs sampler can run in about 5 minutes on a standard laptop (MacBook Pro 2.3 GHz Intel Core i5), and the prediction simulations take about 5 minutes as well. If more nodes/cores were available, prediction simulations can be computed in parallel and further decrease wait time. Calculation of the MLEs and Hessians can be done offline; their combined calculations average around 2 minutes per storm, with a maximum of about 11 minutes. Thus, the total computational time of our approach is the sum of the computational time for the calculation of MLEs and Hessian matrices with the computational time for the Gibbs sampler. Hence, if the computation of the MLEs is performed in parallel as we do here, then the computational time is less than 17 minutes. Given the large computational burden of ensemble prediction systems, which can take hours, we believe this method provides an efficient alternative. The code is available in the Supplementary Material \citep{walsh2022supplementCode}.

\section{Simulation Study}
\label{simstudy}
We present here the results of a simulation study to evaluate the statistical properties of the proposed methods to estimate $\boldsymbol{B}, \boldsymbol{\Sigma_\theta}, \boldsymbol\theta_1, \dots, \boldsymbol\theta_N$. To illustrate adequate coverage of true spatial parameter values based on normal approximations of the MLEs, we simulate error fields with mean zero and random exponential covariance parameters $\boldsymbol\theta_l$. The true $\boldsymbol\theta_l$ values are draws from $\boldsymbol\theta_l \sim N(\tilde{\boldsymbol{B}}\boldsymbol{x}_i, \tilde{\boldsymbol{\Sigma}}_{\boldsymbol\theta})$ where $\tilde{\boldsymbol{B}}$ and $\tilde{\boldsymbol{\Sigma}}_{\boldsymbol\theta}$ are provided in Section \ref{application_to_EFs}. Given that the spatial resolution and size of the spatial domain are critical factors governing the coverage rates, we choose to use the buffer regions $\mathcal{B}_i$ for each of the storms in the training set as well as the corresponding $\boldsymbol{x}_i$ values to generate $\boldsymbol\theta_l$. We simulate error field values for each $\mathcal{B}_i$ and $\boldsymbol{x}_i$ 50 times for a total of 2,350 simulated error fields. 

To create 95\% confidence intervals, we obtain MLEs and approximate the covariance matrix with $\boldsymbol H_l^{-1}$ as described above, with $l \in \{1,\dots,2350\}$. Our results show that we have adequate coverage for both $\theta_{l1}$ and $\theta_{l2}$ in these intervals, with coverage occurring for 95.7\% and 93.1\% of all simulations, respectively. The coverage for $\theta_{l2}$ is lower than 95\% as a result of the likelihood for $\theta_{l2}$ having a slightly heavier right tail which results in a slight decrease in coverage when using the normal approximation. There was very weak correlation between $n_i$ and the corresponding average coverage rate of the 50 simulations, implying that the coverages were not heavily influenced by the number of grid points. 

To illustrate the utility of the Gibbs sampler, we evaluate the credible intervals for $\boldsymbol\theta_1, \dots, \boldsymbol\theta_N,\boldsymbol{B}$ and $\boldsymbol{\Sigma_\theta}$ by looking at the coverage of the true values in each interval. The intervals contain the true $\theta_{l1}$ and $\theta_{l2}$ values in 94.9\% and 92\% of the simulations, respectively. The true generating values for $\boldsymbol{B}$ and $\boldsymbol{\Sigma_\theta}$ are covered 95.7\% and 94\%, respectively. 

\section{Application to Error Fields}
\label{application_to_EFs}

After the data processing (see Section \ref{dataproc_prep}), we interpolate, transform and apply buffers to each of the $N=47$ training storms. We obtain $\boldsymbol{y}_i$ for each TC and calculate $\hat{\boldsymbol\theta}_i$ and $\boldsymbol{H}_i$ for $i \in \{1,\dots,N\}$.  Using the MLEs and Hessian matrices as inputs, we run the Gibbs sampler and analyze the output.

\begin{figure}[t]
    \centering
    \includegraphics[width=5.61893in]{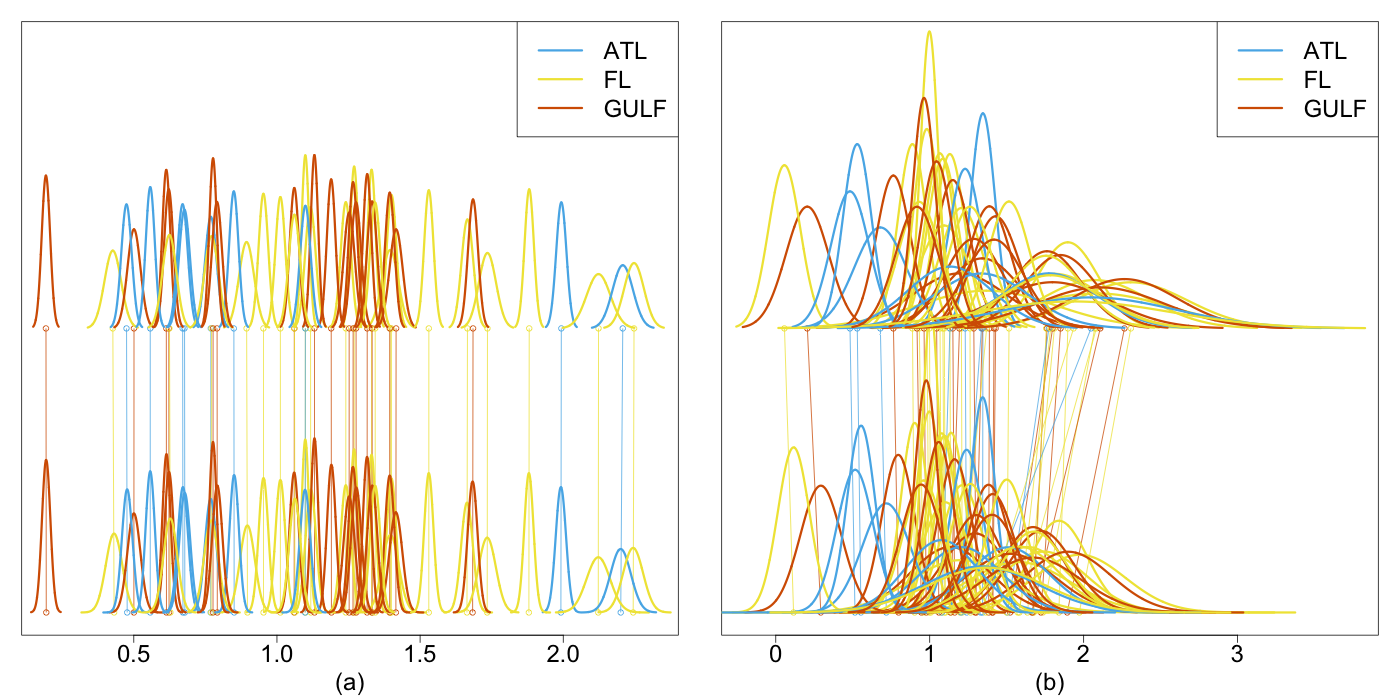}%
    \caption{Distributions of MLEs (top) and for the posterior samples (bottom). (a): $\theta_{i1} = \log(\sigma_i^2 / \phi_i$), $i \in \{1, ..., N\}$; \quad (b) $\theta_{i2} = \log(\sigma_i^2)$, $i \in \{1, ..., N\}$.}
    \label{fig:mleVSposterior_thetas}
\end{figure}

Alongside each of these posterior estimates we have a measure of variance derived from the collection of posterior draws after burn-in. This allows us to compare the variability in our posterior estimates with the variability of the MLEs determined by the asymptotic covariance matrices derived from the Hessians. As illustrated in \autoref{fig:mleVSposterior_thetas}, we see heteroskedasticity in the MLEs, as the estimates' variability tends to increase as the estimates increase for $\theta_{i2}=\log(\sigma_i^2)$. That is, larger values of $\hat{\theta}_{i2} = \log(\hat{\sigma}_i^2)$ tend to have larger variances so the influence of the prior is most noticeable here. MLE estimates for $\theta_{i1}=\log(\sigma_i^2/\phi_i)$ have very high precision and, as a result, the posterior estimates do not show notable differences from the corresponding MLEs. The difference in the precisions of the MLEs of $\theta_{i1}$ and $\theta_{i2}$ is explained by theoretical and simulation results given in \cite{zhang2004inconsistent}, which shows that under in-fill asymptotics both MLEs are asymptotically unbiased but their variances exhibit different behaviors as the sample size increases; while the variance of the MLE of $\theta_{i1}$ converges to zero, the variance of the MLE of $\theta_{i2}$ decreases to a positive value and stops decreasing for larger sample sizes. Additionally, the within-TC uncertainty for each $\boldsymbol{\theta}_i$ (based on the Hessian matrices) differs from that of the between-TC variability. Our hierarchical framework models this variability of parameters between TCs through $\boldsymbol{\Sigma_\theta}$ estimated from the data, which is crucial for generating well-calibrated probabilistic predictions from deterministic forecasts for incoming storms.

The posterior medians for each entry in the regression coefficient matrix $\boldsymbol{B}$ and covariance matrix $\boldsymbol{\Sigma_\theta}$ are:
\[
\tilde{\boldsymbol{B}}=
\left[ \begin{array}{lll}
1.036_{(0.709,1.357)} & 0.258_{(-0.133,0.643)} & 0.017_{(-0.391,0.413)} \\
1.052_{(0.721,1.396)} & 0.229_{( -0.171,0.637)} & 0.217_{(-0.201,0.627)} \\
\end{array} \right]
\]
\[
\tilde{\boldsymbol{\Sigma}}_{\boldsymbol\theta}= \left[ \begin{array}{ll}
0.235_{( 0.160,0.367)}  &  0.064_{(-0.008,0.155)} \\
0.064_{(-0.008,0.155)}  &  0.209_{( 0.129,0.347)} \\
\end{array} \right]
,
\]

\noindent with subscripts indicating the 95\% credible interval for each element. The columns of $\boldsymbol{B}$ represent the Atlantic baseline, the Florida effects and the Gulf effects. Rows correspond to the elements of $\boldsymbol\theta$. The second and third columns of $\boldsymbol{B}$ all contain 0 in the 95\% credible intervals, suggesting the model with a common mean across landfall regions will likely suffice for this dataset.

\subsection{Model Selection}
\label{modelselection}

We use the Laplace-Metropolis estimator to estimate the integrated likelihood for each of three different modeling regimes for $\boldsymbol\theta$ \citep{lewis1997estimating}. We consider three competing models with Model 1 assuming $\boldsymbol\theta_i \sim N(\boldsymbol{Bx}_i, \boldsymbol{\Sigma_\theta})$, with $\boldsymbol{B}$ and $\boldsymbol{x}_i$ specified to model the effects of the Atlantic, Florida and Gulf landfall regions. Model 2 assumes a common mean across all TCs, such that $\boldsymbol\theta_i \sim N(\boldsymbol{\mu_\theta}, \boldsymbol{\Sigma_\theta})$ and Model 3 assumes $\boldsymbol\theta_i = \boldsymbol{\mu_\theta}$, where $\boldsymbol{\mu_\theta}$ is a common mean for $\boldsymbol\theta_i$ across all landfall regions. Note that Model 3 is the only model that drops the hierarchical component of the model corresponding to $\boldsymbol{\Sigma_\theta}$. 

The integrated log-likelihood estimates for Models 1, 2 and 3 are $-75.42$, $-74.49$ and $-13044.51$, respectively. This shows strong support for the hierarchical model framework; under the non-hierarchical Model 3 the uncertainty in the Hessians do not sufficiently explain the variability of the parameters' estimates across different TCs. Model 2 has the most support from the training data; this can be explained by the similarity of average spatial parameter estimates between landfall locations  (see Figure S2 in the Supplementary Material, \cite{walsh2022supplement}) and the penalization for 4 additional hyperparameters to estimate for Model 1. Therefore, we will use Model 2 in the following UQ applications.

\section{Uncertainty Quantification for Future Storms}
\label{uq_future_storms}
With Gibbs sampler output, we validate our method with TCs from 2018 and 2019, using the NAM forecasts to quantify uncertainty related to precipitation that will not yet be observed in a real-time setting. Prior to landfall, we obtain current track forecast data from the NHC's tropical cyclone forecast advisories (\url{https://www.nhc.noaa.gov/archive/}). Details of these advisories and other NHC products can be found at \url{https://www.nhc.noaa.gov/aboutnhcprod.shtml}. For each forecast initialization, a corresponding advisory is reported three hours afterward; we use this information to construct the buffer region for an incoming TC. 

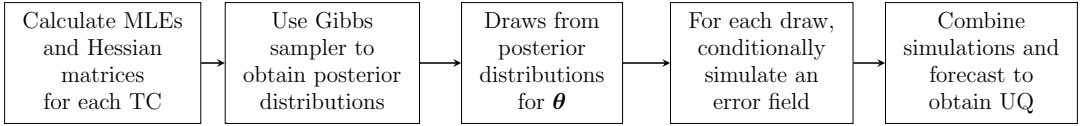
\begin{figure}[t]
    \centering
    \tikzstyle{every node}=[font=\large]
    \tikzstyle{process} = [rectangle, minimum width=2cm, minimum height=1cm, text centered, draw=black]
    \tikzstyle{arrow} = [thick,->,>=stealth]
    
    \noindent\resizebox{5.61893in}{!}{\begin{tikzpicture}[node distance=2cm]
    
    \node (pro1) [process] {\begin{tabular}{c} Calculate MLEs \\ and Hessian \\ matrices \\ for each TC \end{tabular}};
    \node (pro2) [process, right of=pro1, xshift=2cm] {\begin{tabular}{c} Use Gibbs \\ sampler to \\ obtain posterior \\ distributions \end{tabular}};
    \node (pro3) [process, right of=pro2, xshift=2cm] {\begin{tabular}{c} Draws from \\ posterior \\ distributions \\ for $\boldsymbol\theta$  \end{tabular}};
    \node (pro4) [process, right of=pro3, xshift=2cm] {\begin{tabular}{c} For each draw, \\ conditionally \\ simulate an \\ error field \end{tabular}};
    \node (pro5) [process, right of=pro4, xshift=2cm] {\begin{tabular}{c} Combine \\ simulations and \\ forecast to \\ obtain UQ \end{tabular}};
    
    \draw [arrow] (pro1) -- (pro2);
    \draw [arrow] (pro2) -- (pro3);
    \draw [arrow] (pro3) -- (pro4);
    \draw [arrow] (pro4) -- (pro5);
    
    \end{tikzpicture}}
    \caption{Framework to obtain inference and uncertainty quantification results.}
    \label{fig:inference_framework}
\end{figure}

Assume that we have observed $N$ storms and a new storm is coming. Let $\boldsymbol{y}_{N+1}$ be the error field for the $(N+1)$th storm and $\boldsymbol{\theta}_{N+1}$ be the corresponding vector of parameters. Then, according to our hierarchical model, the predictive density of $\boldsymbol{y}_{N+1}$ given the data from the previous storms is

\begin{eqnarray}
    \label{eqn:pred_dens}
    p(\boldsymbol{y}_{N+1}|\boldsymbol{y}_1,\ldots,\boldsymbol{y}_N) &=&
    \int\int\int p(\boldsymbol{y}_{N+1}|\boldsymbol\theta_{N+1})p(\boldsymbol{\theta}_{N+1}|\boldsymbol{\mu_\theta},\boldsymbol{\Sigma_\theta})
    \\ \nonumber 
    && \qquad\qquad\qquad\quad
    p(\boldsymbol{\mu_\theta},\boldsymbol{\Sigma_\theta}|\boldsymbol{y}_1,
    \ldots,\boldsymbol{y}_N) d\boldsymbol{\theta}_{N+1}d\boldsymbol{\mu_\theta}d\boldsymbol{\Sigma_\theta}.
\end{eqnarray}

\begin{figure}[t]
    \centering
    \includegraphics[width=5.61893in]{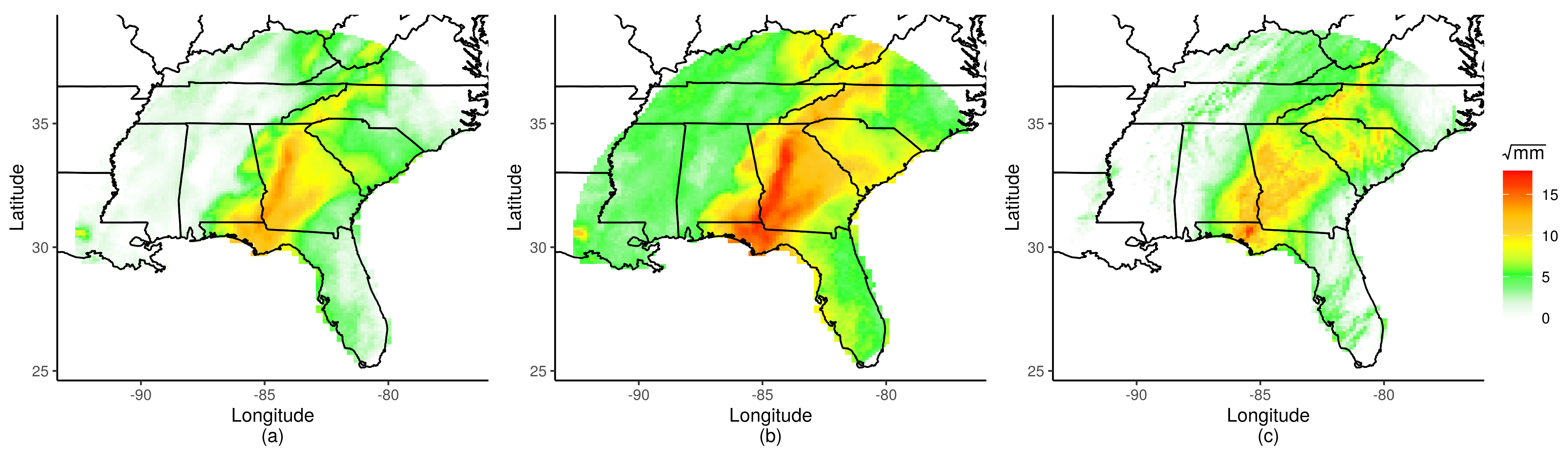}
    \caption{Square root precipitation fields for (a) the NAM forecast, (b) the 95\% prediction map, and (c) the Stage IV data product for Hurricane Michael, 2018.}
    \label{fig:pred_NAM_us_ST4}
\end{figure}

Guided by Equation \eqref{eqn:pred_dens}, we use the output from the MCMC algorithm proposed in Section \ref{computations} to simulate a sample from the predictive distribution of $\boldsymbol{y}_{N+1}$.
Let $(\boldsymbol{\mu_\theta}^{(g)},\boldsymbol{\Sigma_\theta}^{(g)}), g \in \{1,\ldots,G\}$ be the sample from the posterior distribution of $(\boldsymbol{\mu_\theta}, \boldsymbol{\Sigma_\theta})$ obtained with the MCMC algorithm outlined in Section \ref{computations}. We can then simulate a sample from the prior distribution of $\boldsymbol{\theta}_{N+1}|\boldsymbol{y}_1,\ldots,\boldsymbol{y}_N$ using conditional sampling as $\boldsymbol{\theta}^{(g)}_{N+1}=\boldsymbol{\mu_\theta}^{(g)} + \boldsymbol{\omega}^{(g)}$, where $\boldsymbol\omega^{(g)} \sim N(\boldsymbol 0, \boldsymbol{\Sigma_\theta}^{(g)})$. We then generate $\boldsymbol{y}_{N+1}^{(g)}$ from its conditional distribution $\boldsymbol{y}_{N+1}|\boldsymbol{\theta}_{N+1}^{(g)}$. With this, we can generate 1000 corresponding error fields and create 95\% prediction maps by adding the pointwise 95th percentiles of the generated error fields to the available NAM forecast. We can see if the prediction maps are greater than approximately 95\% of grid points for the corresponding Stage IV data. The results for each of the three models are shown in \autoref{tab:cov_pred_storms}. For Model 2, the average of coverages for the 95\% prediction maps for the six TCs in the test set is about 96.86\% and the 99\% prediction maps show average coverages of 98.81\%. These maps of precipitation totals illustrate potential worst-case scenarios based on the uncertainty in the forecast (see \autoref{fig:pred_NAM_us_ST4}b). We also look at coverages based on the extreme low and high forecasted precipitation. We define low values to be below 2.5mm (0.098 in) of precipitation and above the 95th percentile for each particular storm, respectively. Subsetting by these thresholds can show the coverages for these extremes. For the low extremes, 95th and 99th upper bound coverages of Model 2 are 98.87\% and 99.85\%, and for high extremes we have 95.55\% and 97.35\% coverage.

\indent For the six prediction storms, Florence and Dorian were Atlantic storms, Alberto and Michael were Florida storms, and Gordon and Barry were Gulf storms. Thus, if we compare coverage rates by landfall location, we obtain the results shown in the bottom of \autoref{tab:cov_pred_storms}. Models 1 and 2 assume $\boldsymbol\theta_i$ follow a normal distribution, where landfall region is and is not considered in the mean, respectively. From the table, we see approximately the equal average coverages, regardless of whether a location-specific mean is used; this further supports the idea that Model 2 is the best model (which agrees with the results from the Laplace-Metropolis estimators of the integrated likelihoods).

\begin{table}[h]
\begin{center}
 \begin{tabular}{|l |c c | c c| c c|}  
 \hline 
 \multicolumn{1}{|c|}{} & \multicolumn{2}{c|}{Model 1} & \multicolumn{2}{c|}{Model 2}&\multicolumn{2}{c|}{Model 3}\\
\cline{2-7}
  Storm & 95\% & 99\% & 95\% & 99\% & 95\% & 99\% \\ [0.5ex] 
  \hline\hline
Alberto, 2018 & 0.9753 & 0.9951 & 0.9768 & 0.9960 & 0.9651 & 0.9890\\
\hline
Florence, 2018 & 0.9356 & 0.9651 & 0.9492 & 0.9715 & 0.9356 & 0.9610\\
\hline
Gordon, 2018 & 0.9616 & 0.9816 & 0.9633 & 0.9821 & 0.9537 & 0.9737\\
\hline
Michael, 2018 & 0.9421 & 0.9810 & 0.9478 & 0.9837 & 0.9224 & 0.9676\\
\hline
Barry, 2019 & 0.9832 & 0.9958 & 0.9840 & 0.9954 & 0.9772 & 0.9897\\
\hline
Dorian, 2019 & 0.9967 & 1.0000 & 0.9991 & 1.0000 & 0.9937 & 1.0000\\
\hline\hline
Atlantic Average & 0.9613 & 0.9798 & 0.9701 & 0.9835 & 0.9600 & 0.9774\\
\hline
Florida Average & 0.9607 & 0.9889 & 0.9641 & 0.9906 & 0.9463 & 0.9796\\
\hline
Gulf Average & 0.9716 & 0.9882 & 0.9730 & 0.9883 & 0.9647 & 0.9811\\
\hline\hline
Overall Average & 0.9647 & 0.9865 & 0.9687 & 0.9881 & 0.9561 & 0.9796\\
 \hline
\end{tabular}
\caption{\label{tab:cov_pred_storms} Coverage rates for the Stage IV precipitation of the six TCs in the test dataset for the 95\% and 99\% prediction maps based on 1000 error fields added to the corresponding NAM forecast. Models 1, 2, and 3 are defined as in Section \ref{modelselection}. Each of the three landfall regions contained two prediction storms.}
\end{center}
\end{table}

The prediction maps provide precipitation totals for a given percentile. Alternatively, if one is interested in the probability of precipitation surpassing a particular threshold (e.g., 2 inches) at different locations, then a probability map can be produced. Adding an error field to the NAM forecast will create one potential realization for the observed precipitation; an indicator map can be constructed to check whether or not a given grid point has surpassed the given threshold. Repeating this 1000 times and aggregating indicator plots will provide probabilistic information regarding which locations are most likely to experience severe rain.

One approach to evaluating our prediction intervals is comparing their widths with the observed forecast errors in the test storms. Obtaining the 95\% prediction interval at each grid point for each prediction storm, the average length of these intervals is 1.606 inches, which indicates a margin of error of 0.803in. If we look at the 95\% upper bound of these 95\% prediction intervals, the margin of error is 2.161in. The maximum margin of error from every 95\% prediction interval over all prediction storms was 6.607in. We note that the 95\% upper bound, 99\% upper bound and maximum absolute difference between the forecasted values (from NAM) and observed values (from Stage IV) in the prediction storms were 1.54in, 3.93in and 16.109in respectively. For the grid points with the largest forecasted precipitation totals in the 24 hour window, we obtain margins of error that are approximately half of this total (see Figure S11 in the Supplementary Material \cite{walsh2022supplement}). Therefore, we find our prediction intervals to be reasonable and useful.

\begin{figure}[H]
    \centering
    \includegraphics[width=5.61893in]{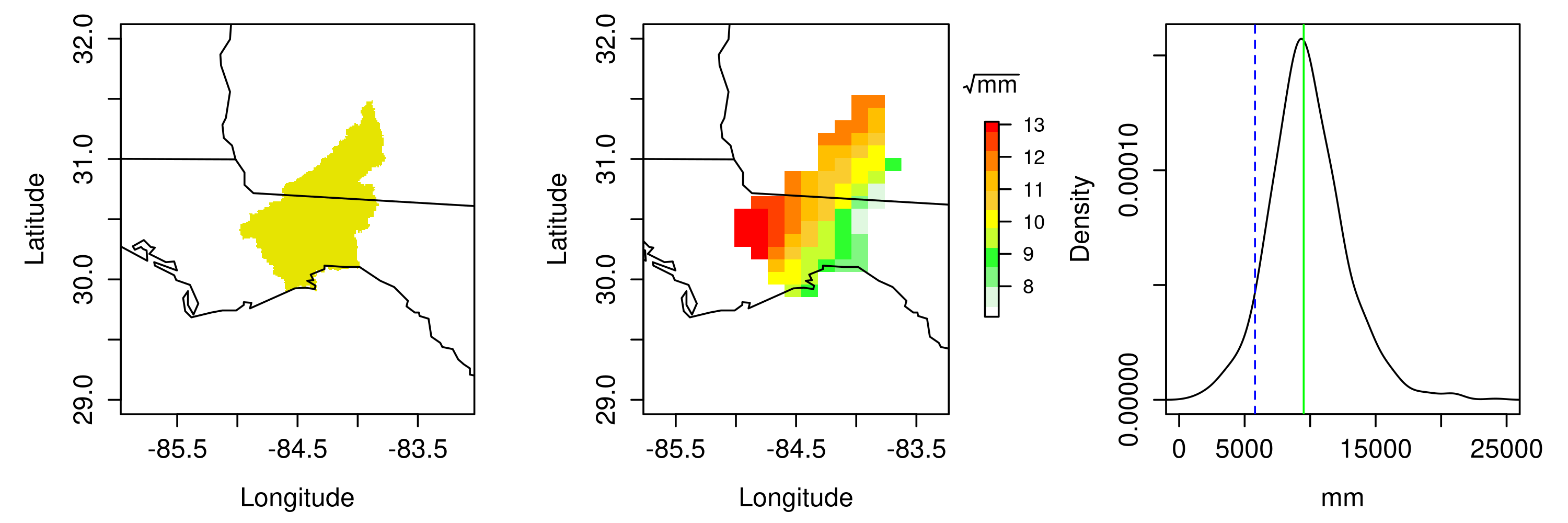}
    \caption{(a) The Ochlockonee subregion, (b) the interpolated Ochlockonee subregion showing 24 hour accumulated precipitation (in square root mm) from Hurricane Michael, and (c) the predictive density for the accumulated precipitation (in mm) for the subregion in the 24 hour time window. The solid line indicates the total from the NAM forecast and the dashed line indicates the total from the Stage IV data.}
    \label{fig:hydroapp}
\end{figure}

To illustrate another application of our method, we study the variability of a TC's accumulated precipitation for a hydrologic subregion. The United States Geological Survey defines hydrologic regions, subregions and other areas based upon the drainage locations for CONUS. We select the Ochlockonee hydrologic subregion as an example since it is contained within the buffer region of Hurricane Michael and it also contains the city of Tallahassee (see \autoref{fig:hydroapp}). Using the aforementioned prediction output, we transform the potential precipitation realizations to millimeters and aggregate over the hydrologic subregion. From this, a predictive distribution for the accumulated precipitation for the subregion can be obtained, for example by using kernel density estimation. This distribution can be further transformed to obtain potential volumes of water that will pass through the subregion's rivers as a result of the TC. In the example for Hurricane Michael, we can see that the predictive distribution successfully captures the observed value based upon Stage IV data (see \autoref{fig:hydroapp}c).

\subsection{Scores for competing UQ approaches}
\label{scores_for_UQ_log}
The Ochlockonee subregion described above is one of 90 different watersheds we study to evaluate our proposed methodology. Across the six prediction storms, we consider watersheds located across seven states (Alabama, Florida, Georgia, Lousiana, Mississippi, North Carolina and South Carolina). Over 100 of these watersheds are contained within one or more of the six prediction storms' buffer regions. We set a requirement that at least 30 grid points of the buffer region must be contained in a watershed to qualify, and we proceed by analyzing TC precipitation  for the six prediction storms at these 90 watersheds that qualify.
\newline\indent
To assess the performance of our proposed method, we compare performance with multiple candidate models. We consider multiple approaches for modeling the spatial parameters $\boldsymbol\theta$, including the three models previously mentioned in \autoref{modelselection}. Additionally, we include a nonparametric approach for modeling $\boldsymbol\theta$, Model 4, where a bootstrap sample is drawn from the 47 MLEs of the training storms. Model 5 does not consider spatial dependence between grid points, and thus reduces $\boldsymbol\theta$ to a scalar $\theta=\log(\sigma^2)$. In Models 1-5, there is no bias adjustment performed (i.e., $\boldsymbol\mu=\boldsymbol 0$). Models 6-9 will correspond to Models 1-4 except that we consider a bias adjustment such that the pointwise mean has been subtracted from each NAM forecast (see Appendix B in the Supplementary Material, \cite{walsh2022supplement}).
\newline\indent
We use the logarithmic scoring rule to evaluate the predictive densities from each of the nine candidate models for the 90 watersheds \citep{good1952rational, gneiting2007strictly}. The logarithmic score has many desirable properties: it is a proper scoring rule that takes spatial dependence into account and, because it is equivalent to the log of the predictive density, it is related to posterior model probabilities and Bayes factors.

\begin{table}[h]
\centering
\begin{tabular}[t]{|l|r|r|r|r|r|r|r|r|r|}
\hline
  Model & 1 &  2 &  3 &  4 &  5 &  6 &  7 &  8 &  9\\
  \hline
   \multicolumn{1}{|l|}{sample $\boldsymbol\theta$} &
   \multicolumn{1}{c|}{$N$, $\boldsymbol{Bx}_i$} &
   \multicolumn{1}{c|}{$N$, $\boldsymbol{\mu_\theta}$} &
   \multicolumn{1}{c|}{\raisebox{-0.9pt}{$\bar{\boldsymbol\theta}$}, fixed} &
   \multicolumn{1}{c|}{boot} & 
   \multicolumn{1}{c|}{nonsp} & 
   \multicolumn{1}{c|}{$N$, $\boldsymbol{Bx}_i$} &
   \multicolumn{1}{c|}{$N$, $\boldsymbol{\mu_\theta}$} &
   \multicolumn{1}{c|}{\raisebox{-0.9pt}{$\bar{\boldsymbol\theta}$}, fixed} &   \multicolumn{1}{c|}{boot} \\
        \hline
      \multicolumn{1}{|l|}{bias adjust} & \multicolumn{5}{c|}{No pointwise bias adjustment} & 
   \multicolumn{4}{c|}{Bias adjustment by $\boldsymbol\mu$}\\
\hline\hline
log score & -12506 & \textbf{-12313} & -19026 & -12851 &-24824&-13333 & -13303 & -19349 & -13470   \\
\hline
\end{tabular}
\caption{\label{tab:sum_log_scores} For each of the nine candidate models, the sum of the log predictive densities over the 90 prediction basins from the 6 prediction storms is presented. Models 1 and 2 use a hierarchical model with a Gaussian distribution for $\boldsymbol\theta$, where the mean does ($\boldsymbol{Bx}_i$) and does not ($\boldsymbol{\mu_\theta}$) vary by landfall location ($\boldsymbol x_i$), respectively. Model 3 is similar to Model 2, but removes modeling of $\boldsymbol{\Sigma_\theta}$, so the only variability of $\boldsymbol\theta$ is from the asymptotic precision matrices derived from the Hessians. Model 4 uses a bootstrap sample of the MLEs $\hat{\boldsymbol\theta}$. Model 5 is a nonspatial version of Model 1, which reduces $\boldsymbol\theta$ to a scalar, $\log(\sigma^2)$. Models 6-9 are equivalent to Models 1-4, respectively, except for a bias adjustment using the posterior mean of $\boldsymbol\mu$ (Appendix C). The best score is shown in boldface.}
\end{table}

\begin{table}[h]
\centering
\begin{tabular}[t]{|l|r|r|r|r|r|r|r|r|r|}
\hline
  Model & 1 &  2 &  3 &  4 &  5 &  6 &  7 &  8 &  9\\
  \hline
   \multicolumn{1}{|l|}{sample $\boldsymbol\theta$} &
   \multicolumn{1}{c|}{$N$, $\boldsymbol{Bx}_i$} &
   \multicolumn{1}{c|}{$N$, $\boldsymbol{\mu_\theta}$} &
   \multicolumn{1}{c|}{\raisebox{-0.9pt}{$\bar{\boldsymbol\theta}$}, fixed} &
   \multicolumn{1}{c|}{boot} & 
   \multicolumn{1}{c|}{nonsp} & 
   \multicolumn{1}{c|}{$N$, $\boldsymbol{Bx}_i$} &
   \multicolumn{1}{c|}{$N$, $\boldsymbol{\mu_\theta}$} &
   \multicolumn{1}{c|}{\raisebox{-0.9pt}{$\bar{\boldsymbol\theta}$}, fixed} &   \multicolumn{1}{c|}{boot} \\
        \hline
      \multicolumn{1}{|l|}{bias adjust} & \multicolumn{5}{c|}{No pointwise bias adjustment} & 
   \multicolumn{4}{c|}{Bias adjustment by $\boldsymbol\mu$}\\
\hline\hline
ATL &      \textbf{16}&\textbf{16}&3& 7&0&0&0&0&1\\\hline
FL &       \textbf{10}& 2&2&1 &0&0&0&0&1\\\hline
GULF &     7& \textbf{14}&1& 9&0&0&0&0&0\\\hline\hline
noncoastal&\textbf{13}& 11&2&7&0&0&0&0&0\\\hline
coastal &  20&\textbf{21}&4&10&0&0&0&0&2\\\hline\hline
overall best&\textbf{33}&32&6&17&0&0&0&0&2\\\hline
\end{tabular}
\caption{\label{tab:which_scores_best} For each of the nine candidate models, the number of times the model had the highest score for each of the prediction basins. These results are subset by landfall region (selected by a majority vote of grid points), and also if the basin was coastal or not. The best performer for each category is set in boldface.}
\end{table}

When we sum the log predictive densities from each watershed, we obtain a total score for each candidate model. These sums are found in \autoref{tab:sum_log_scores}. Here, we see that Model 2, a hierarchical model with a common mean $\boldsymbol{\mu_\theta}$ for each $\boldsymbol\theta$ and no bias adjustment, has the best performance. When we consider each basin individually, we see that Models 1 and 2 (both hierarchical models with normal distributions for $\boldsymbol\theta$ and no bias adjustment) each are selected as the best about the same number of times (see \autoref{tab:which_scores_best}). If we subset basins by the three landfall regions or by whether the basin is coastal or not, this persists, although it seems Florida basins prefer Model 1 and Gulf basins generally prefer Model 2. So, we don't have a definite answer as to whether or not different watersheds should have different distributions for the model parameters. It is worth noting that even when we subset basins by landfall region or their coastal status, the best model overall by the log score is still Model 2 (\autoref{tab:which_scores_best}). This is likely due to the fact that Model 2 generally has slightly wider tails than Model 1, so it is not as heavily penalized when there is a larger error between the forecasted and observed values. Model 4 is outperformed by Models 1 and 2 providing evidence that, in this application, the Bayesian approach outperforms the use of MLEs alone.

One of the most appealing aspects of Bayesian methods is the natural updating process as new data becomes available. After an incoming storm is processed by the UQ algorithm and the Stage IV products become available, MLEs can be calculated for the storm and the posterior distributions from the Gibbs sampler will be updated accordingly. As the number of storms increase, we obtain more information with which to quantify the uncertainty for future storms. Although the updating process could be done by simple weighting (linear combination of the old and current forecasts), we prefer our approach because the updates are based on the data (via $\boldsymbol\theta$) and hence will not depend on some chosen vector of weights as in a simple weighting scheme.

\section{Discussion}

Many NWP models produce precipitation forecasts with limited information regarding their uncertainty. When dealing with extreme events like TCs, it is important to quantify this uncertainty to better inform those in the storm's path. With our Bayesian framework, we propose a novel approach to analyze the variability in NWP forecast errors and provide UQ for these extreme weather events with a limited number of storms. Our approach maintains spatial coherence and allows modeling of these rare events where previous postprocessing methods would deteriorate. 

We study the operational NAM TC forecasts and obtain the corresponding QPEs available from Stage IV. We also learn about the amount of uncertainty we can expect from a given NAM forecast for TC precipitation through estimation of $\boldsymbol\theta_i$ and the implementation of our UQ algorithm. This general framework can be implemented with other NWP models (eg: Global Forecast System or Hurricane Weather Research and Forecast models) to explore these results and also compare their performances with that from the NAM. 

This work inspires avenues for future research. The effects of using alternative, or even nonparametric, distributions for $\boldsymbol\theta$ or $\boldsymbol y_i$ should be explored. Additionally, there is uncertainty not only in the amount of precipitation, but also in the track that the eye of the TC follows. Including uncertainty within the storm track can help improve our results by having less dependence on the particular NWP model forecast being studied. From here, additional uncertainty with respect to the landfall region of the storm could be pursued as well. We plan to create a more complex model by introducing nonstationarity for each error field, as well as a spatiotemporal component to allow for changes in the structure of the error fields over time, e.g., the subsequent 24 hour window of time for each TC.

Our methodology can be expanded to ensembles. Also, by incorporating our UQ methodology within the Bayesian model averaging or nonhomogeneous Gaussian regression frameworks \citep{raftery2005using, gneiting2005calibrated}, or adapting the \cite{kennedy2001bayesian} framework, this approach may illuminate the types of uncertainties not generally characterized in most ensemble approaches.

In conclusion, we provide a novel framework for quantifying the uncertainty of tropical storm and hurricane forecasts. This technique can help to illuminate systematic biases in forecasts as well as better understand the variability within a particular forecast. The framework can also be applied to other spatial fields, such as wind forecasts. By implementing the UQ algorithm, one is able to have a clearer understanding of the potential variability in a TC precipitation forecast. With greater understanding of the abilities and limits of a QPF, we hope to inform research scientists of new approaches for assessing accuracy and reliability of their products.

\begin{center}
{\large\bf Acknowledgments}
\end{center}The authors would like to thank the anonymous referees, the Associate Editor and the Editor for their conscientious efforts and constructive comments, which improved the quality of this paper.

\begin{center}
{\large\bf SUPPLEMENTARY MATERIAL}
\end{center}

Supplement to ``A Bayesian Hierarchical Model Framework to Quantify Uncertainty of Tropical Cyclone Precipitation Forecasts''
(DOI:10.1214/[provided by typesetter]; .pdf) The supplementary material contains Appendices A, B, C, and D. Appendix A contains derivations for the full conditional distributions used in the Gibbs sampler and analytical calculations for the Hessian matrices of the parameters of the exponential covariance function. Appendix B provides details on modeling the systematic bias of the NAM forecast. Appendix C contains simulation study results, and Appendix D provides plots for each of the tropical cyclone landfalls within the training set and 95\% upper bounds for uncertainty of each TC in the test set. Additionally, plots of the margins of error for each test storm, and an example log-likelihood surface for the original and reparameterized parameter space are provided.

{Supplement to ``A Bayesian Hierarchical Model Framework to Quantify Uncertainty of Tropical Cyclone Precipitation Forecasts''} 
{Code for ``A Bayesian Hierarchical Model Framework to Quantify Uncertainty of Tropical Cyclone Precipitation Forecasts''} 
(DOI:10.1214/[provided by typesetter]; .zip) This file contains the code and some processed data to reproduce the results from the manuscript. It is also available at \url{https://github.com/stevewalsh124/NAM-Model-Validation}.

\bibliographystyle{jasa} 
\bibliography{bibfile} 

\newpage
\begin{center}
{\large\bf SUPPLEMENTARY MATERIAL}
\end{center}
Supplementary Material for A Bayesian Hierarchical Model Framework to Quantify Uncertainty of Tropical Cyclone Precipitation Forecasts

\begin{appendix}

\renewcommand{\thefigure}{S\arabic{figure}}
\setcounter{figure}{0}

\section{Derivations}
\subsection{Full Conditionals for Gibbs Sampler}
\label{apdx:full_condls}

\noindent Below is the joint posterior distribution from which the full conditionals are derived. Recall the prior choices of $\pi(\boldsymbol{B})\propto 1$ and $\pi(\boldsymbol{\Sigma_\theta})\equiv IW(\nu_0,\boldsymbol{S_0})$.
\begin{align*}
    \pi(\boldsymbol{\theta}_i, \boldsymbol{B}, \boldsymbol{\Sigma_\theta}|\boldsymbol{y}_i^*, \boldsymbol{x_i})
    &=\pi(\boldsymbol{\theta}_i, \boldsymbol{B}, \boldsymbol{\Sigma_\theta}|\boldsymbol{\hat\theta}_i, \boldsymbol{x_i}) \\
    &= \mathcal{L}(\boldsymbol{\hat\theta}_i|\boldsymbol{\theta}_i,\boldsymbol{H}_i^{-1})\times
    \pi(\boldsymbol{\theta}_i|\boldsymbol{\boldsymbol{Bx_i},\Sigma_\theta})\times
    \pi(\boldsymbol{B})\times
    \pi(\boldsymbol{\Sigma_\theta})\\
    &\propto \bigg\{\prod_{i=1}^{N} \exp\Big(-\frac{1}{2}(\boldsymbol{\hat{\theta}_i}-\boldsymbol{\theta_i})^T\boldsymbol{H_i}(\boldsymbol{\hat{\theta}_i}-\boldsymbol{\theta_i})\Big)\bigg\}
    \\
    &\qquad\times\bigg\{\prod_{i=1}^{N} |\boldsymbol{\Sigma_\theta}|^{-1/2}\exp\Big(-\frac{1}{2}(\boldsymbol{\theta_i}-\boldsymbol{Bx_i})^T \boldsymbol{\Sigma_\theta}^{-1}(\boldsymbol{\theta_i-Bx_i})\Big)\bigg\}\\
    &\qquad\times1\times|\boldsymbol{\Sigma_\theta}|^{-(\nu_0+p+1)/2}\exp(-\text{tr}(\boldsymbol{S_0\Sigma_\theta^{-1}})/2)
\end{align*}

We begin by obtaining the full conditional for $\boldsymbol\theta_i$.
\begin{align*}
    \pi(\boldsymbol\theta_i|-)&\propto \bigg\{\prod_{i=1}^{N} \exp\Big(-\frac{1}{2}(\boldsymbol{\hat{\theta}_i}-\boldsymbol{\theta_i})^T\boldsymbol{H_i}(\boldsymbol{\hat{\theta}_i}-\boldsymbol{\theta_i})\Big)\bigg\}
    \\
    &\qquad \times
    \bigg\{\prod_{i=1}^{N} |\boldsymbol{\Sigma_\theta}|^{-1/2}\exp\Big(-\frac{1}{2}(\boldsymbol{\theta_i}-\boldsymbol{Bx}_i)^T \boldsymbol{\Sigma_\theta}^{-1}(\boldsymbol{\theta_i-Bx}_i)\Big)\bigg\}\\
    &\propto \exp\bigg(-\frac{1}{2}\sum_{i=1}^N\big[(\hat{\boldsymbol\theta}_i-\boldsymbol\theta_i)^T\boldsymbol{H}_i(\boldsymbol{\hat{\theta}_i}-\boldsymbol{\theta_i})+ (\boldsymbol{\theta_i}-\boldsymbol{Bx}_i)^T \boldsymbol{\Sigma_\theta}^{-1}(\boldsymbol{\theta_i-Bx}_i)\big]\bigg)\\
    & \propto \exp\bigg(-\frac{1}{2}\sum_{i=1}^N \big[ \boldsymbol{\theta}_i^T(\boldsymbol{H_i}+\boldsymbol{\Sigma_\theta}^{-1})\boldsymbol\theta_i  - 2\boldsymbol\theta_i^T(\boldsymbol{H_i\hat{\theta}_i}+\boldsymbol{\Sigma_\theta}^{-1}\boldsymbol{Bx}_i)\big]\bigg)\\
    & \equiv N\Big((\boldsymbol{H_i}+\boldsymbol{\Sigma_\theta}^{-1})^{-1}(\boldsymbol{H_i\hat{\theta}_i}+\boldsymbol{\Sigma_\theta}^{-1}\boldsymbol{Bx}_i),(\boldsymbol{H_i}+\boldsymbol{\Sigma_\theta}^{-1})^{-1}\Big)
\end{align*}

Next we find the full conditional for $\boldsymbol{\Sigma_\theta}$, which is a $p \times p$ matrix, with $p$ being the number of elements contained in the vector of parameters $\boldsymbol\theta$.
\begin{align*}
    \pi(\boldsymbol{\Sigma_\theta}|-)
    &\propto \bigg\{\prod_{i=1}^{N} |\boldsymbol{\Sigma_\theta}|^{-\frac{1}{2}}\exp\Big(-\frac{1}{2}(\boldsymbol{\theta_i}-\boldsymbol{Bx}_i)^T \boldsymbol{\Sigma_\theta}^{-1}(\boldsymbol{\theta_i-Bx}_i)\Big)\bigg\}\\
    &\qquad \times|\boldsymbol{\Sigma_\theta}|^{-\frac{\nu_0+p+1}{2}}\exp(-\text{tr}(\boldsymbol{S_0\Sigma_\theta^{-1}})/2)\\
    &\propto \bigg\{|\boldsymbol{\Sigma_\theta}|^{-\frac{N}{2}}\exp\Big(-\frac{1}{2}\sum_{i=1}^{N}(\boldsymbol{\theta_i}-\boldsymbol{Bx}_i)^T \boldsymbol{\Sigma_\theta}^{-1}(\boldsymbol{\theta_i-Bx}_i)\Big)\bigg\}\\
    &\qquad \times|\boldsymbol{\Sigma_\theta}|^{-\frac{\nu_0+p+1}{2}}\exp(-\text{tr}(\boldsymbol{S_0\Sigma_\theta^{-1}})/2)\\
    &\propto \bigg\{|\boldsymbol{\Sigma_\theta}|^{-\frac{N}{2}}\exp\Big(-\frac{1}{2}\text{tr}\big(\sum_{i=1}^{N}(\boldsymbol{\theta_i}-\boldsymbol{Bx}_i)^T \boldsymbol{\Sigma_\theta}^{-1}(\boldsymbol{\theta_i-Bx}_i)\big)\Big)\bigg\}\\
    &\qquad \times|\boldsymbol{\Sigma_\theta}|^{-\frac{\nu_0+p+1}{2}}\exp(-\text{tr}(\boldsymbol{S_0\Sigma_\theta^{-1}})/2)\\
    &\propto \bigg\{|\boldsymbol{\Sigma_\theta}|^{-\frac{N}{2}}\exp\Big(-\frac{1}{2}\text{tr}\Big(\sum_{i=1}^{N}(\boldsymbol{\theta_i-Bx}_i)(\boldsymbol{\theta_i}-\boldsymbol{Bx}_i)^T \boldsymbol{\Sigma_\theta}^{-1}\Big)\Big)\bigg\}\\
    &\qquad \times|\boldsymbol{\Sigma_\theta}|^{-\frac{\nu_0+p+1}{2}}\exp(-\text{tr}(\boldsymbol{S_0\Sigma_\theta^{-1}})/2)\\
    &\propto |\boldsymbol{\Sigma_\theta}|^{-(N+\nu_0+p+1)/2}
    \exp\Big(-\frac{1}{2}\text{tr}\Big(\big[\sum_{i=1}^{N}(\boldsymbol{\theta_i-Bx}_i)(\boldsymbol{\theta_i}-\boldsymbol{Bx}_i)^T+\boldsymbol{S}_0\big] \boldsymbol{\Sigma_\theta}^{-1}\Big)\Big)\\
    &\equiv IW\Big(N+\nu_0,\sum_{i=1}^N (\boldsymbol\theta_i-\boldsymbol{Bx}_i)(\boldsymbol\theta_i-\boldsymbol{Bx}_i)^T+\boldsymbol{S}_0\Big)
\end{align*}

So, the full conditional of $\boldsymbol{\Sigma_\theta}$ is distributed inverse Wishart.

The following equation is the probability density function of the matrix-variate normal distribution:

$p(\textbf{X}|\textbf{M},\textbf{U},\textbf{V}) = \frac{\exp\Big(-\frac{1}{2}\text{tr}\big[\textbf{V}^{-1}(\textbf{X}-\textbf{M})^T\textbf{U}^{-1}(\textbf{X}-\textbf{M})\big]\Big)}{(2\pi)^{np/2}|\textbf{V}|^{n/2}|\textbf{U}|^{p/2}}$

Note $\textbf{X} \in \mathbb{R}^{n\times p}$ is the random variable, $\textbf{M}$ is the mean matrix, and $\textbf{U}\in \mathbb{R}^{n\times n}$ and $\textbf{V}\in \mathbb{R}^{p\times p}$ represent the among-row and among-column variance, respectively.
Expanding the kernel of the matrix-variate distribution we find the following equivalent expressions that will help determine the full conditional for $\boldsymbol{B}$.
\begin{align*}
    p(\textbf{X}|\textbf{M},\textbf{U},\textbf{V}) &\propto \exp\Big(-\frac{1}{2}\text{tr}\big[\textbf{V}^{-1}(\textbf{X}-\textbf{M})^T\textbf{U}^{-1}(\textbf{X}-\textbf{M})\big]\Big)\\
    &\propto \exp\Big(-\frac{1}{2}\text{tr}\big[\textbf{V}^{-1}(\textbf{X}^T\textbf{U}^{-1}\textbf{X}-\textbf{X}^T\textbf{U}^{-1}\textbf{M}-\textbf{M}^T\textbf{U}^{-1}\textbf{X}+\textbf{M}^T\textbf{U}^{-1}\textbf{M})\big]\Big)\\    
    &\propto \exp\Big(-\frac{1}{2}\text{tr}\big[\textbf{V}^{-1}\textbf{X}^T\textbf{U}^{-1}\textbf{X}-\textbf{V}^{-1}\textbf{X}^T\textbf{U}^{-1}\textbf{M} -\\
    &\qquad\qquad\qquad\qquad\textbf{V}^{-1}\textbf{M}^T\textbf{U}^{-1}\textbf{X}+\textbf{V}^{-1}\textbf{M}^T\textbf{U}^{-1}\textbf{M}\big]\Big)\\
\end{align*}

Let us obtain the full conditional for $\boldsymbol{B}$:
\begin{align*}
    \pi(\boldsymbol{B}|-) &\propto \prod_{i=1}^{N} \exp\Big(-\frac{1}{2}(\boldsymbol{\theta_i}-\boldsymbol{Bx_i})^T \boldsymbol{\Sigma_\theta}^{-1}(\boldsymbol{\theta_i}-\boldsymbol{Bx_i})\Big)\\
    &\propto \exp\Big(-\frac{1}{2}\sum_{i=1}^N\big[\boldsymbol{x_i^TB^T\Sigma_\theta^{-1}Bx_i}-\boldsymbol{\theta_i^T\Sigma_\theta^{-1}Bx_i}-\boldsymbol{x_i^TB^T\Sigma_\theta^{-1}\theta_i} \big] \Big)\\
    &\propto \exp\Big(-\frac{1}{2}\sum_{i=1}^N\text{tr}\big[\boldsymbol{x_i^TB^T\Sigma_\theta^{-1}Bx_i}-\boldsymbol{\theta_i^T\Sigma_\theta^{-1}Bx_i}-\boldsymbol{x_i^TB^T\Sigma_\theta^{-1}\theta_i} \big] \Big)\\
    &\propto \exp\Big(-\frac{1}{2}\Big(\text{tr}\big[\sum_{i=1}^N\boldsymbol{x_i^TB^T\Sigma_\theta^{-1}Bx_i}\big]-\text{tr}\big[\sum_{i=1}^N\boldsymbol{\theta_i^T\Sigma_\theta^{-1}Bx_i}\big]-\\
    &\qquad\qquad\qquad\qquad\text{tr}\big[\sum_{i=1}^N\boldsymbol{x_i^TB^T\Sigma_\theta^{-1}\theta_i} \big] \Big)\Big)\\
    &\propto \exp\Big(-\frac{1}{2}\Big(\text{tr}\big[\sum_{i=1}^N(\boldsymbol{x_ix_i^T})\boldsymbol{B^T\Sigma_\theta^{-1}B}\big]-\text{tr}\big[\sum_{i=1}^N(\boldsymbol{x_i\theta_i^T})\boldsymbol{\Sigma_\theta^{-1}B}\big]-\\
    &\qquad\qquad\qquad\qquad\text{tr}\big[\sum_{i=1}^N\boldsymbol{x_i^TB^T\Sigma_\theta^{-1}\theta_i} \big] \Big)\Big)
\end{align*}
So from the first portion of the expression we obtain $\textbf{V}^{-1}=\sum_{i=1}^N\boldsymbol{x}_i\boldsymbol{x}_i^T$ and $\textbf{U}^{-1}=\boldsymbol{\Sigma_\theta}^{-1}$. The second and third show that $\textbf{MV}^{-1}=\sum_{i=1}^N(\boldsymbol{\theta}_i\boldsymbol{x}_i^T)$, which implies that $\textbf{M} = (\sum_{i=1}^N\boldsymbol{\theta}_i\boldsymbol{x}_i^T)(\sum_{i=1}^N \boldsymbol{x}_i\boldsymbol{x}_i^T)^{-1}$. Therefore, we have the full conditional for the matrix $\boldsymbol{B}$:

$\pi(\boldsymbol{B}|-) \equiv MN\Big((\sum_{i=1}^N\boldsymbol{\theta}_i\boldsymbol{x}_i^T)(\sum_{i=1}^N \boldsymbol{x}_i\boldsymbol{x}_i^T)^{-1}, \boldsymbol{\Sigma_\theta}, (\sum_{i=1}^N\boldsymbol{x}_i\boldsymbol{x}_i^T)^{-1}\Big)$

\subsection{Analytical Hessian calculations for Exponential Covariance Function}
\label{supp:hess_calcs}

To obtain a Hessian matrix based on the MLEs of $\boldsymbol\theta = ([\log(\sigma^2/\phi), \log(\sigma^2)]^T$, we can use the multivariate delta method \citep{lehmanncasella} after calculating the MLEs and corresponding Hessian matrix with respect to $\boldsymbol\lambda=[\sigma^2,\phi]^T$. Let's assume we have a zero-mean spatial process with no nugget and exponential covariance function with marginal variance $\sigma^2$ and scale parameter $\phi$, that is $\Sigma(\phi)_{i,j}=\Sigma(\phi,Y)_{i,j}=\exp(-\frac{||Y_i-Y_j||}{\phi})$ with $||Y_i-Y_j||$ represents the Euclidean distance between the $i$th and $j$th observations. Then, if we have a spatially correlated $n-$vector $Y_n \equiv Y \sim N(0, \sigma^2\Sigma(\phi))$, we can express the likelihood in the following manner:

\begin{align*}
    \mathscr{L} = \mathscr{L}(\phi, \sigma^2|Y)&=(2\pi\sigma^2)^{-n/2}|\Sigma(\phi)|^{-1/2}\exp\Big(-\frac{1}{2\sigma^2}Y^T\Sigma^{-1}(\phi)Y\Big)\\
    l = l(\phi,\sigma^2|Y)&=-\frac{n}{2}\log(2\pi)-\frac{n}{2}\log(\sigma^2)-\frac{1}{2}\log|\Sigma(\phi)|-\frac{1}{2\sigma^2}Y^T\Sigma^{-1}(\phi)Y 
\end{align*}

Taking the derivative twice with respect to $\sigma^2$, we get the following:

\begin{align*}
    \frac{\partial l}{\partial \sigma^2} &= \frac{-n}{2\sigma^2}+\frac{1}{2(\sigma^2)^2}Y^T\Sigma^{-1}(\phi)Y\\
    \frac{\partial^2 l}{(\partial \sigma^2)^2} &= \frac{n}{2(\sigma^2)^2}-\frac{1}{(\sigma^2)^3}Y^T\Sigma^{-1}(\phi)Y
    \approx \frac{n}{2(\sigma^2)^2}-\frac{n}{(\sigma^2)^2}=-\frac{n}{2(\sigma^2)^2}
\end{align*}

Before we look at the derivatives with respect to $\phi$, let's consider some important rules that we can use:

\begin{align*}
    \frac{\partial \Sigma^{-1}(\phi)}{\partial \phi} &= -\Sigma^{-1}(\phi) \frac{\partial \Sigma(\phi)}{\partial \phi}\Sigma^{-1}(\phi)\\
    \frac{\partial \log|\Sigma(\phi)|}{\partial\phi} &= tr\bigg[\Sigma^{-1}(\phi)\frac{\partial \Sigma(\phi)}{\partial \phi}\bigg]
\end{align*}

We can see that each of these require $\frac{\partial \Sigma(\phi)}{\partial \phi} =\Sigma'(\phi)= \Sigma(\phi) \odot \frac{D}{\phi^2}$, where $\odot$ represents the Hadamard (element-wise) product of two $m \times n$ matrices (in this case $m=n$), $D$ is the $n \times n$ matrix of (Euclidean) distances between all of the $N$ locations where $Y$ is observed \citep[e.g.,][Chapter 5]{gramacy2020surrogates}. Let us begin with the derivatives with respect to $\phi$:
 
  \begin{align*}
     \frac{\partial^2 l}{\partial \phi \partial \sigma^2} &= \frac{\partial}{\partial\phi}\bigg[\frac{-n}{2\sigma^2}+\frac{1}{2(\sigma^2)^2}Y^T\Sigma^{-1}(\phi)Y\bigg]\\
     &= \frac{1}{2(\sigma^2)^2}Y^T\frac{\partial}{\partial\phi}\big[\Sigma^{-1}(\phi)\big]Y\\
     &= -\frac{1}{2(\sigma^2)^2}Y^T\Sigma^{-1}(\phi)\Sigma'(\phi)\Sigma^{-1}(\phi)Y
 \end{align*}
 
 Now we can start working to obtain the final derivative which will be used to calculate the Hessian, $\frac{\partial^2 l}{\partial\phi^2}:$
 
 \begin{align*}
     \frac{\partial l}{\partial \phi} &= \frac{\partial}{\partial\phi}\bigg[-\frac{1}{2}\log|\Sigma(\phi)|-\frac{1}{2\sigma^2}Y^T\Sigma^{-1}(\phi)Y\bigg]\\
     &= -\frac{1}{2}tr\Big[\Sigma^{-1}(\phi)\Sigma'(\phi)\Big] + \frac{1}{2\sigma^2}Y^T\Sigma^{-1}(\phi)\Sigma'(\phi)\Sigma^{-1}(\phi)Y
 \end{align*}
  
 Recall $\frac{\partial \Sigma(\phi)}{\partial \phi} =\Sigma'(\phi)= \Sigma(\phi) \odot \frac{D}{\phi^2}$. Let's start the second derivative:
 
 \begin{align*}
     \frac{\partial^2l}{\partial\phi^2} &= \frac{\partial}{\partial\phi}\bigg[-\frac{1}{2}tr\Big[\Sigma^{-1}(\phi)\Sigma'(\phi)\Big] + \frac{1}{2\sigma^2}Y^T\Sigma^{-1}(\phi)\Sigma'(\phi)\Sigma^{-1}(\phi)Y\bigg]\\
     &= -\frac{1}{2}tr(A) + \frac{1}{2\sigma^2}Y^TBY
 \end{align*}
 
 Here, 
 
 \begin{align*}
     A &= \frac{\partial}{\partial\phi}\Big[\Sigma^{-1}(\phi)\Sigma'(\phi)\Big]\\
     B &= \frac{\partial}{\partial\phi}\Big[\Sigma^{-1}(\phi)\Sigma'(\phi)\Sigma^{-1}(\phi)\Big]
 \end{align*}
 
 Before finding more specific expressions for A and B, let's take a moment to find the second derivative of $\Sigma(\phi)$, that is $\frac{\partial^2\Sigma(\phi)}{\partial\phi^2}$. We need to use $\partial(X\odot Y)=\partial(X)\odot Y+X\odot\partial(Y)$ \citep{matrixcookbook}:
 
 \begin{align*}
     \Sigma''(\phi) = \frac{\partial^2\Sigma(\phi)}{\partial\phi} = \frac{\partial}{\partial\phi}\Sigma'(\phi) &= \frac{\partial}{\partial\phi}\Big[\Sigma(\phi) \odot \frac{D}{\phi^2}\Big]\\
     &= \Sigma'(\phi) \odot \frac{D}{\phi^2} - \Sigma(\phi) \odot \frac{2D}{\phi^3} 
 \end{align*}
 
 So, let's evaluate A:
 
  \begin{align*}
     A &= \frac{\partial}{\partial\phi}\Big[\Sigma^{-1}(\phi)\Sigma'(\phi)\Big]\\
     &= -\Sigma^{-1}(\phi)\Sigma'(\phi)\Sigma^{-1}(\phi)\Sigma'(\phi) + \Sigma^{-1}(\phi)\Sigma''(\phi)\\     
     &= -\bigg(\Sigma^{-1}(\phi)\Sigma'(\phi)\bigg)^2 + \Sigma^{-1}(\phi)\Sigma''(\phi)
 \end{align*}

Here we evaluate B:

\begin{align*}
         B &= \frac{\partial}{\partial\phi}\Big[\Sigma^{-1}(\phi)\Sigma'(\phi)\Sigma^{-1}(\phi)\Big]\\
         &= \frac{\partial}{\partial\phi}\Big[\Sigma^{-1}(\phi)\Big]\Sigma'(\phi)\Sigma^{-1}(\phi) +
         \Sigma^{-1}(\phi)\frac{\partial}{\partial\phi}\Big[\Sigma'(\phi)\Big]\Sigma^{-1}(\phi) +
         \Sigma^{-1}(\phi)\Sigma'(\phi)\frac{\partial}{\partial\phi}\Big[\Sigma^{-1}(\phi)\Big]\\
         &= \Big[-\Sigma^{-1}(\phi)\Sigma'(\phi)\Sigma^{-1}(\phi)\Big]\Sigma'(\phi)\Sigma^{-1}(\phi) +
         \Sigma^{-1}(\phi)\Big[\Sigma''(\phi)\Big]\Sigma^{-1}(\phi) \\ &\qquad -
         \Sigma^{-1}(\phi)\Sigma'(\phi)\Big[\Sigma^{-1}(\phi)\Sigma'(\phi)\Sigma^{-1}(\phi)\Big]\\
         &= \Sigma^{-1}(\phi)\Big[\Sigma''(\phi)\Big]\Sigma^{-1}(\phi) -2
         \bigg(\Sigma^{-1}(\phi)\Sigma'(\phi)\bigg)^2\Sigma^{-1}(\phi)\\
         &= \Sigma^{-1}(\phi)\Big[\Sigma''(\phi)\Big]\Sigma^{-1}(\phi) -2
         \Sigma^{-1}(\phi)\Sigma'(\phi)\Sigma^{-1}(\phi)\Sigma'(\phi)\Sigma^{-1}(\phi)
\end{align*}

With our second derivatives in hand, we can take the expectation of each of these expressions to simplify them. This is using the result from Appendix A of \cite{berger2001objective}, where it states that if $X \sim N(\mu, \Sigma)$ and $A$ is a symmetric matrix, then $\mathbb{E}(X^TAX)=tr(A\Sigma)+\mu^TA\mu$.

So, that will give us the following:

\begin{align*}
    \mathbb{E}\bigg(\frac{\partial^2 l}{(\partial\sigma^2)^2}\bigg) &= \mathbb{E}\bigg[\frac{n}{2(\sigma^2)^2}-\frac{1}{(\sigma^2)^3}Y^T\Sigma^{-1}(\phi)Y\bigg] \\
    &= \frac{n}{2(\sigma^2)^2}-\frac{1}{(\sigma^2)^3}\mathbb{E}\bigg[Y^T\Sigma^{-1}(\phi)Y\bigg] \\ 
    &= \frac{n}{2(\sigma^2)^2}-\frac{1}{(\sigma^2)^3}tr\Big(\sigma^2\Sigma^{-1}(\phi)\Sigma(\phi)\Big) + 0^T\Sigma^{-1}(\phi)0\\ 
    &= \frac{n}{2(\sigma^2)^2}-\frac{n}{(\sigma^2)^2}\\
    &= \frac{-n}{2(\sigma^2)^2}\\
\end{align*}

Since we have that $\mu=0$, we can always ignore the $\mu^T A\mu$ term.

\begin{align*}
    \mathbb{E}\bigg(\frac{\partial^2 l}{\partial\sigma^2\partial\phi}\bigg) &= \mathbb{E}\bigg[-\frac{1}{2(\sigma^2)^2}Y^T\Sigma^{-1}(\phi)\Sigma'(\phi)\Sigma^{-1}(\phi)Y\bigg]\\
    &= -\frac{1}{2(\sigma^2)^2}\mathbb{E}\bigg[Y^T\Sigma^{-1}(\phi)\Sigma'(\phi)\Sigma^{-1}(\phi)Y\bigg]\\
    &= -\frac{1}{2(\sigma^2)^2} tr\Big(\sigma^2\Sigma^{-1}(\phi)\Sigma'(\phi)\Sigma^{-1}(\phi)\Sigma(\phi)\Big)\\
    &= -\frac{1}{2\sigma^2} tr\Big(\Sigma^{-1}(\phi)\Sigma'(\phi)\Big)\\
\end{align*}

In the following calculation, note that neither $A$ nor $B$ depend on $Y$.

\begin{align*}
    \mathbb{E}\bigg(\frac{\partial^2 l}{\partial\phi^2}\bigg) &= \mathbb{E}\bigg[-\frac{1}{2}tr(A) + \frac{1}{2\sigma^2}Y^TBY\bigg]\\
    &= -\frac{1}{2}tr(A) + \frac{1}{2\sigma^2}\mathbb{E}\bigg[Y^TBY\bigg]\\
    &= -\frac{1}{2}tr(A) + \frac{1}{2\sigma^2}tr\bigg(\sigma^2 B\Sigma(\phi)\bigg)\\
    &= -\frac{1}{2}tr(A) + \frac{1}{2}tr\bigg(\Big[\Sigma^{-1}(\phi)\Sigma''(\phi)\Sigma^{-1}(\phi) -2\big(\Sigma^{-1}(\phi)\Sigma'(\phi)\big)^2\Sigma^{-1}(\phi)\Big]\Sigma(\phi)\bigg)\\
    &= \frac{1}{2}tr\Big(\big(\Sigma^{-1}(\phi)\Sigma'(\phi)\big)^2 - \Sigma^{-1}(\phi)\Sigma''(\phi)\Big) +\\
    &\qquad\qquad\qquad\qquad\frac{1}{2}tr\Big(\Sigma^{-1}(\phi)\Sigma''(\phi) - 2\Sigma^{-1}(\phi)\Sigma'(\phi)\Sigma^{-1}(\phi)\Sigma'(\phi)\Big)\\
    &= \frac{1}{2}tr\Big(\big(\Sigma^{-1}(\phi)\Sigma'(\phi)\big)^2 - \Sigma^{-1}(\phi)\Sigma''(\phi)+ \Sigma^{-1}(\phi)\Sigma''(\phi) -\\
    &\qquad\qquad\qquad\qquad2\Sigma^{-1}(\phi)\Sigma'(\phi)\Sigma^{-1}(\phi)\Sigma'(\phi)\Big)\\
    &= \frac{1}{2}tr\Big(\big(\Sigma^{-1}(\phi)\Sigma'(\phi)\big)^2 - 2\Sigma^{-1}(\phi)\Sigma'(\phi)\Sigma^{-1}(\phi)\Sigma'(\phi)\Big)\\
    &= -\frac{1}{2}tr\Big(\big(\Sigma^{-1}(\phi)\Sigma'(\phi)\big)^2\Big)
\end{align*}

\section{Modeling the systematic bias of the NAM forecast}
\label{apdx:mu_bias}
Within our hierarchical framework, we specified a flat prior for $\boldsymbol\mu$ by specifying a precision matrix of zeros: $\boldsymbol C^{-1}=0_{n_\mathcal{D} \times n_\mathcal{D}}$. This setting allows the data to speak for itself. We also employed an informative prior, with a prior mean of $\boldsymbol 0$ and a covariance matrix modeled with an exponential covariance function, such that its prior spatial parameters were estimated by maximum likelihood from the grid points of $\hat{\boldsymbol\mu}$ which had data for 20 or more storms. Upon implementing both of these priors for $\boldsymbol\mu$ and obtaining the corresponding posterior distributions, both were outperformed by specifying no systematic bias, i.e. $\boldsymbol\mu=\boldsymbol 0$.

Therefore, 
We begin by obtaining an estimate $\hat{\boldsymbol\mu}$ of the mean process to approximate the bias-adjusted $\boldsymbol y_i^*$ for $i \in \{1,\dots,N\}$. This is achieved through an empirical average over all error fields: $\hat{\boldsymbol\mu}(\boldsymbol{s})=(1/n_{\mathcal{M}(\boldsymbol{s})})\Sigma_{i \in \mathcal{M}(\boldsymbol{s})} \boldsymbol{y}_i(\boldsymbol{s})$, where $\boldsymbol{s}$ is a grid point location in CONUS, $\mathcal{M}(\boldsymbol{s})$ is the set of storms for which the buffer region contains $\boldsymbol{s}$, and $n_{\mathcal{M}(\boldsymbol{s})}$ is the cardinality of $\mathcal{M}(\boldsymbol{s})$. This allows us to approximate a zero-mean Gaussian process with $\hat{\boldsymbol{y}}_i^*=\boldsymbol{y}_i-\boldsymbol A_i \hat{\boldsymbol\mu}\overset{a}{\sim} N(\boldsymbol{0},\boldsymbol A_i\boldsymbol\Sigma (\boldsymbol \theta_i)\boldsymbol A_i^T)$. For computational reasons, we choose to treat $\hat{\boldsymbol\mu}$ as constant here to avoid simulating $\boldsymbol\mu$ within the MCMC loop.

We also use each $\hat{\boldsymbol\theta}_i$ to learn about the variability of the mean process $\boldsymbol\mu$. For each error field, we estimate the covariance matrix $\boldsymbol\Sigma_i \equiv \boldsymbol{A}_i\boldsymbol\Sigma(\boldsymbol{\hat\theta}_i)\boldsymbol{A}_i^T$ detailing the correlation between grid points for a particular bias-adjusted error field. The $N$ precision matrices $\boldsymbol\Sigma_i^{-1}$ along with the prior precision $\boldsymbol C^{-1}$ allow us to generate $\boldsymbol{\Sigma_\mu}$, the posterior covariance of the mean process $\boldsymbol\mu$:

\begin{equation}
        \boldsymbol{\Sigma_\mu} = \Big[\boldsymbol C^{-1} + \sum_{i=1}^{N} \boldsymbol\Sigma_i^{-1} \Big]^{-1}.
\end{equation}

\noindent This additionally informs $\boldsymbol{m_\mu}$, the posterior mean for $\boldsymbol\mu$:

\begin{equation}
    \boldsymbol{m_\mu}=\boldsymbol0_{n_\mathcal{D}} + \boldsymbol{\Sigma_\mu} \sum_{i=1}^{N}\bigg[\boldsymbol\Sigma_i^{-1} \boldsymbol y_i\bigg].
\end{equation}

Given the inherent interdependence between calculating $\hat{\boldsymbol\theta}_i, i \in \{1,\dots,N\}$ and $\boldsymbol{m_\mu}$, we implement an algorithm with the style of Expectation Maximization. The goal of this algorithm is to jointly optimize both the spatial parameter estimates and the mean process in a balanced and equitable manner. We begin by estimating $\hat{\boldsymbol\mu}$, the MLEs $\hat{\boldsymbol\theta}_i$ and finally $\boldsymbol{m_\mu}$ as described above. We replace our empirical estimate $\hat{\boldsymbol\mu}$ with the posterior mean $\boldsymbol{m_\mu}$ and use this to redefine our bias-adjusted error fields: $\hat{\boldsymbol{y}}_i^*=\boldsymbol y_i - \boldsymbol A_i \boldsymbol{m_\mu}$. Upon redefining $\hat{\boldsymbol{y}}_i^*$ we update our MLEs $\hat{\boldsymbol\theta}_i$ and then update $\boldsymbol{m_\mu}$ similarly. Repeating this process until convergence allows for joint optimization between both $\hat{\boldsymbol\theta}_i, i \in \{1,\dots,N\}$ and $\boldsymbol{m_\mu}$.

\begin{figure}%
    \centering
    \includegraphics[width=5.61893in]{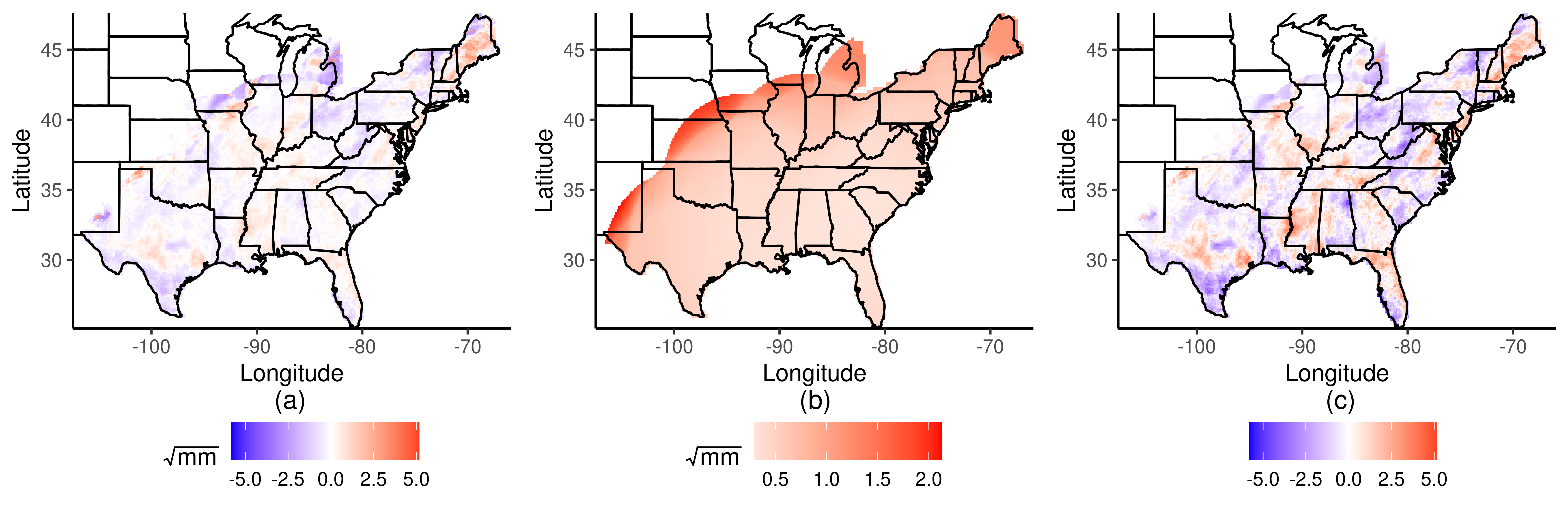}%
    \caption{(a) $\boldsymbol{m_\mu}$, the posterior mean of $\boldsymbol\mu$ estimated from all 47 error fields, (b) posterior standard deviations for each grid point of $\boldsymbol\mu$, and (c) standardized error map based on all 47 error fields. Grid points in (a) with absolute values greater than 5 were set to 5 for to aid comparison with (c); these two plots have a common legend derived from the range of (c).}
    \label{fig:pointwise_plots}
\end{figure}

With estimate $\hat{\boldsymbol\mu}$, we obtain $\hat{\boldsymbol{y}}_i^*$ and calculate $\hat{\boldsymbol\theta}_i$ for $i \in \{1,\dots,N\}$. We implement calculations for $\boldsymbol{m_\mu}$, the posterior mean of the mean process $\boldsymbol\mu$ as described in Section 2.3 (see \autoref{fig:pointwise_plots}a). Upon comparing $\hat{\boldsymbol\mu}$ and $\boldsymbol{m_\mu}$, we find that the two are identical down to 12 decimal places, indicating that we obtain convergence between $\boldsymbol{m_\mu}$ and $\hat{\boldsymbol\theta}_i$ after one iteration when we use an objective prior on $\boldsymbol\mu$ as described in Section 2.2 of the paper. 

We can illustrate the uncertainty for each grid point in $\boldsymbol{m_\mu}$ by creating a map with the corresponding posterior standard deviations, available as the square root of the diagonal elements of $\boldsymbol{\Sigma_\mu}$ (\autoref{fig:pointwise_plots}b). Using $\boldsymbol{m_\mu}$ and $\boldsymbol{\Sigma_\mu}$, we calculate the standardized mean map of errors, as shown in \autoref{fig:pointwise_plots}c. These plots suggest there may be locations where the NAM is systematically biased, with notable areas of overestimation along the Appalachian mountains and southern Texas. We can also see that while southwestern Florida is typically overestimated, there is a tendency for the NAM to underestimate TC precipitation in northeastern Florida.

We illuminate potentially systematic biases of NAM with respect to TC precipitation by producing a posterior mean map for the error fields with $\boldsymbol{m_\mu}$. Given the results of the model comparisons in subsection 5.1 of the paper, models that do not account for systematic bias performed better in this application. More complex models, such as systematic bias that changes through time, are conceptually possible but may create substantial computational burden. More complex models for systematic bias offer an avenue for future research.

\section{Simulation Study Results}
\label{apdx:simstudy}

\begin{figure}%
    \centering
    {{\includegraphics[width=5.5in]{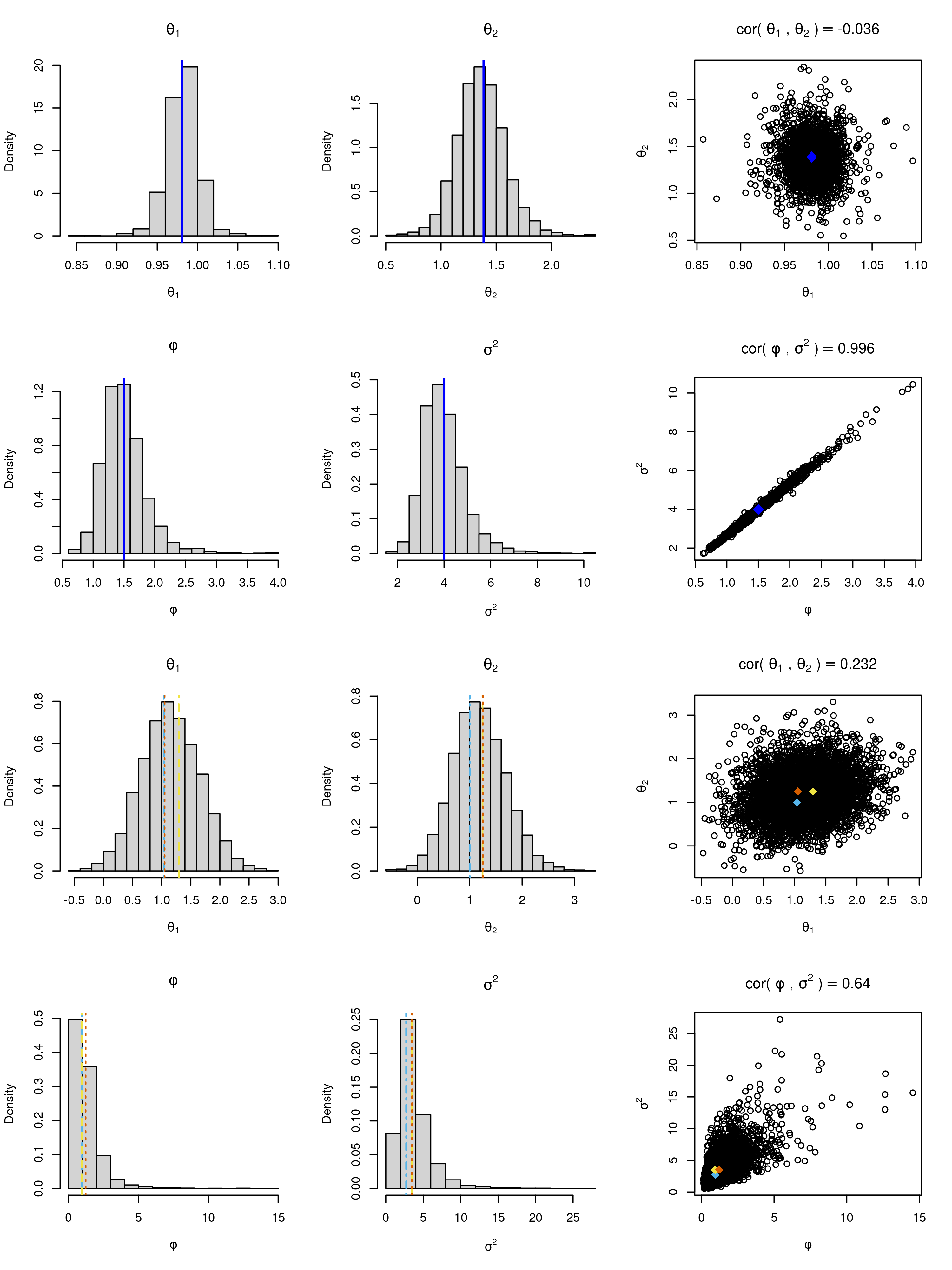} }}%
    \qquad
    \caption{Parameter estimates for 2,350 simulated error fields for the reparameterization $\boldsymbol\theta$ (first row) as well as the original parameterization $\boldsymbol\lambda$ (second row) when true values are fixed at $\sigma^2 = 4, \phi = 1.5$ for each simulation (shown by the blue lines and diamonds). Parameter estimates for 2,350 simulated error fields for the reparameterization $\boldsymbol\theta$ (third row) as well as the original parameterization $\boldsymbol\lambda$ (fourth row) when true values are generated randomly where $\boldsymbol\theta_l \sim N(\tilde{\boldsymbol{B}} \boldsymbol{x}_i,\tilde{\boldsymbol{\Sigma}}_{\boldsymbol\theta})$ shown in Section 4 of the paper. Blue, red and yellow lines and diamonds represent the true parameter values generated by values of Atlantic, Florida and Gulf landfall regions, respectively.}%
    \label{fig:myMLEsim_results}%
\end{figure}

To illustrate adequate coverage of true spatial parameter values based on normal approximations of the MLEs, we simulate error fields with mean zero and exponential covariance parameters of $\sigma^2=4$ and $\phi=1.5$. These true values for the simulations are chosen as they are similar to the average MLE values over all of the error fields in the test set (4.14 and 1.42, repsectively). Upon reparameterizing, this is equivalent to a true parameter vector of $\boldsymbol\theta= [\log(\sigma^2/\phi), \log(\sigma^2)]^T = (0.981,1.386)^T$. In the second row of \autoref{fig:myMLEsim_results}, histograms for $\sigma^2$ and $\phi$ both show a right skew. There is also evidence for a ridge in the likelihood surface of $\boldsymbol\lambda$ \citep[as described by][]{zhang2004inconsistent} with a correlation of 0.996 between $\sigma^2$ and $\phi$. Conversely, the histograms in the first row show the components of $\boldsymbol\theta$ are both approximately normal and the scatter plot shows approximate independence between the two elements of $\boldsymbol\theta$.

In the last two rows of \autoref{fig:myMLEsim_results}, the true $\boldsymbol\theta_l$ values are random samples from $\boldsymbol\theta_l \sim N(\tilde{\boldsymbol{B}}\boldsymbol{x}_i, \tilde{\boldsymbol{\Sigma}}_{\boldsymbol\theta})$ where $\tilde{\boldsymbol{B}}$ and $\tilde{\boldsymbol{\Sigma}}_{\boldsymbol\theta}$ are provided in Section 4 of the paper. Accounting for the variation of estimates across the different TCs, we see that the correlation of the MLEs for $\sigma^2$ and $\phi$ decreases but is still relatively strong at 0.64. The correlation for $\theta_{l1}$ and $\theta_{l2}$ increases but is still relatively weak at 0.232.

\newpage
\section{Plots}
\subsection{Plots of Tropical Cyclone Landfalls and Other Model Output}
\label{apdx: TCplots}

\begin{figure}[H]%
    \centering
    {{\includegraphics[width=5.61893in]{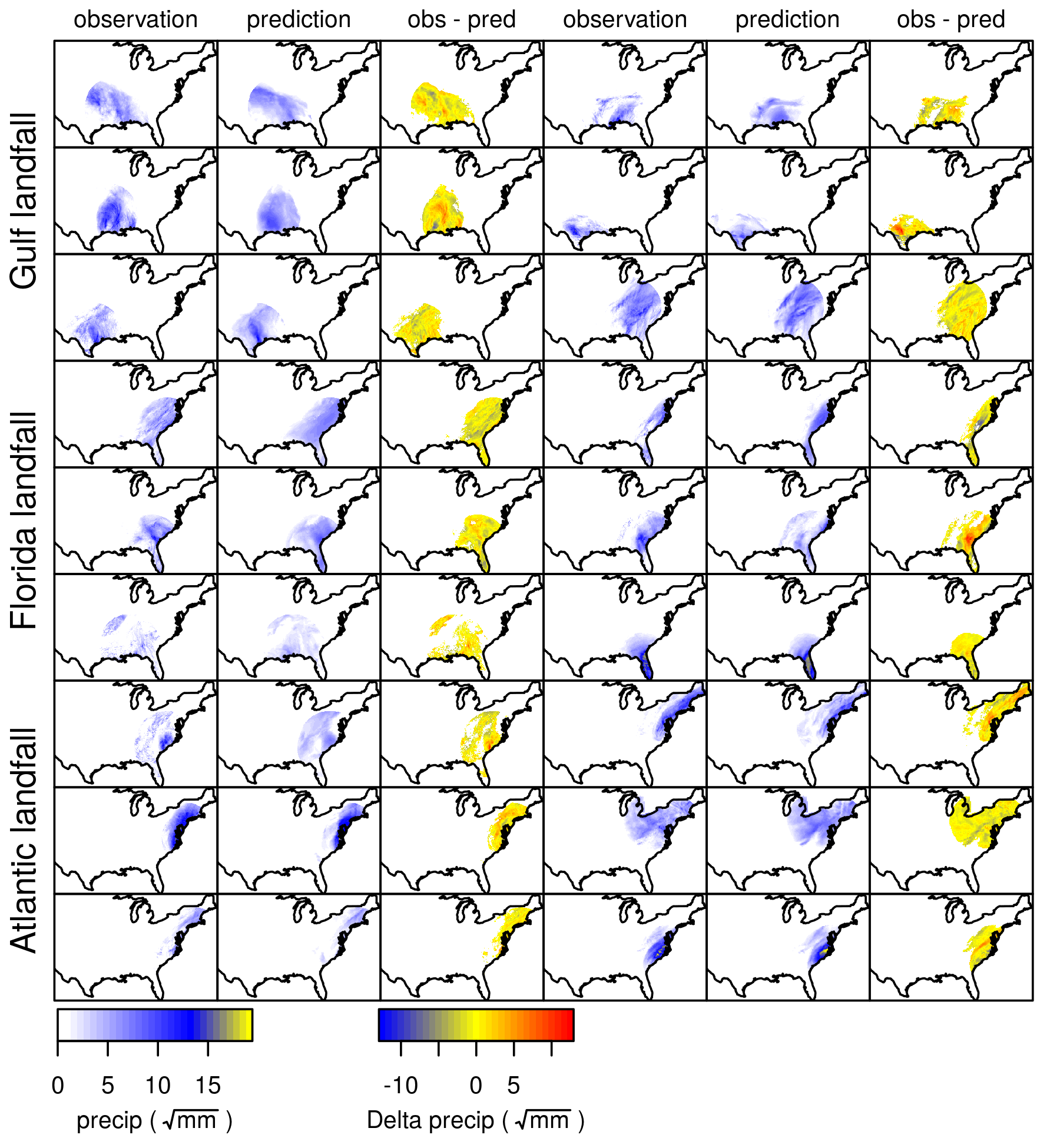} }}%
    \qquad
    \caption{Plots of landfalls for a subset of the 47 training TCs, where 6 of each are selected from each landfall region. The first column represents 24 hour accumulated square root precipitation of the Stage IV data, the second column is the corresponding NAM forecast, and the third column is the difference between Stage IV and NAM. The remaining columns follow this pattern.}%
    \label{fig:someStorms}%
\end{figure}

\begin{sidewaysfigure}[H]%
    \centering
    {{\includegraphics[width=8.96814in]{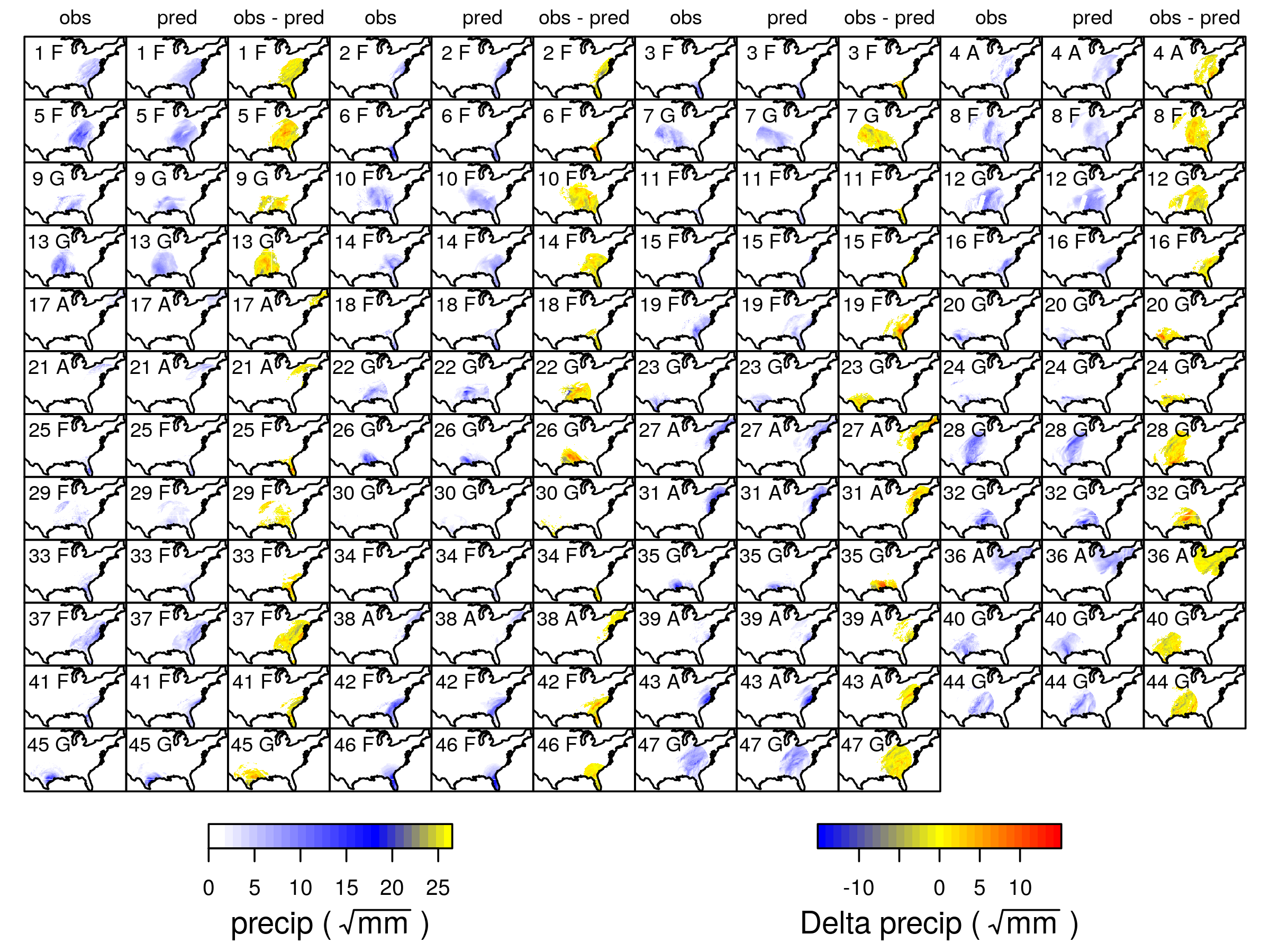} }}%
    \qquad
    \caption{Same as \ref{fig:someStorms}, but for all 47 landfalls for the training TCs. The number and letter in each plot represents the storm number (1-47) and landfall region (A for Atlantic, F for Florida, G for Gulf).}%
    \label{fig:allStorms}%
\end{sidewaysfigure}

\newpage

\subsection{95\% Upper Bounds for Uncertainty of Prediction Storms}
\label{supp:95UB_maps}

These images show the NAM forecast (left), the 95\% upper bound from our UQ method (middle), and the observed precipitation based on Stage IV (right) for each of the six prediction storms from 2018-2019.

\begin{figure}[ht]
    \centering
    \includegraphics[width=5.61893in]{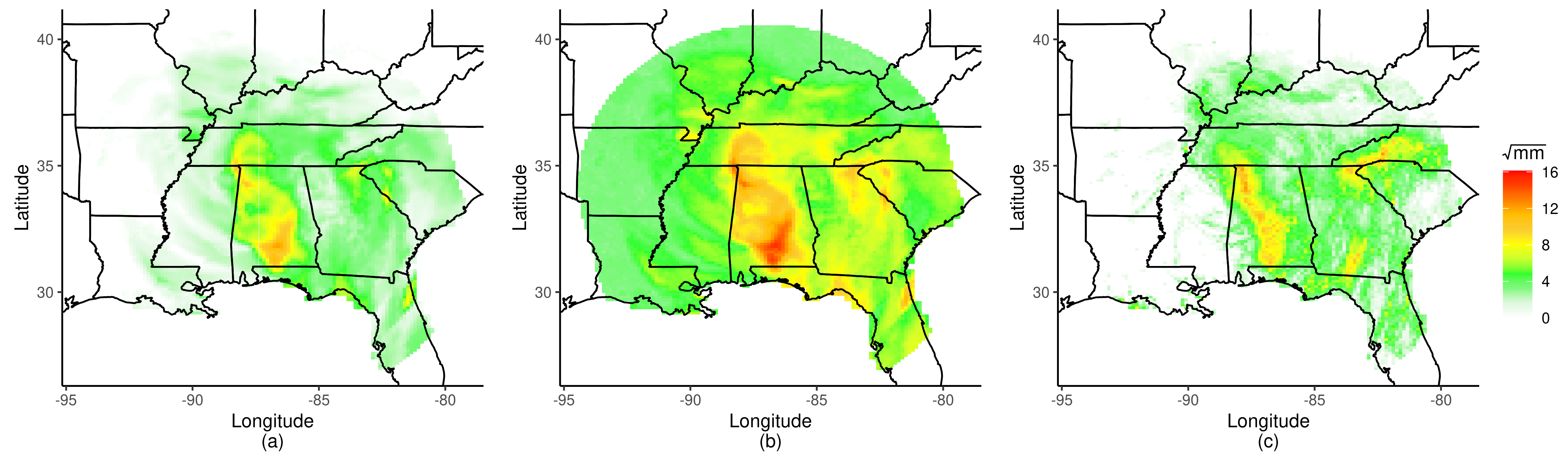}
    \caption{ First prediction storm: Alberto 2018.}
\end{figure}
\begin{figure}[ht]
    \centering
    \includegraphics[width=5.61893in]{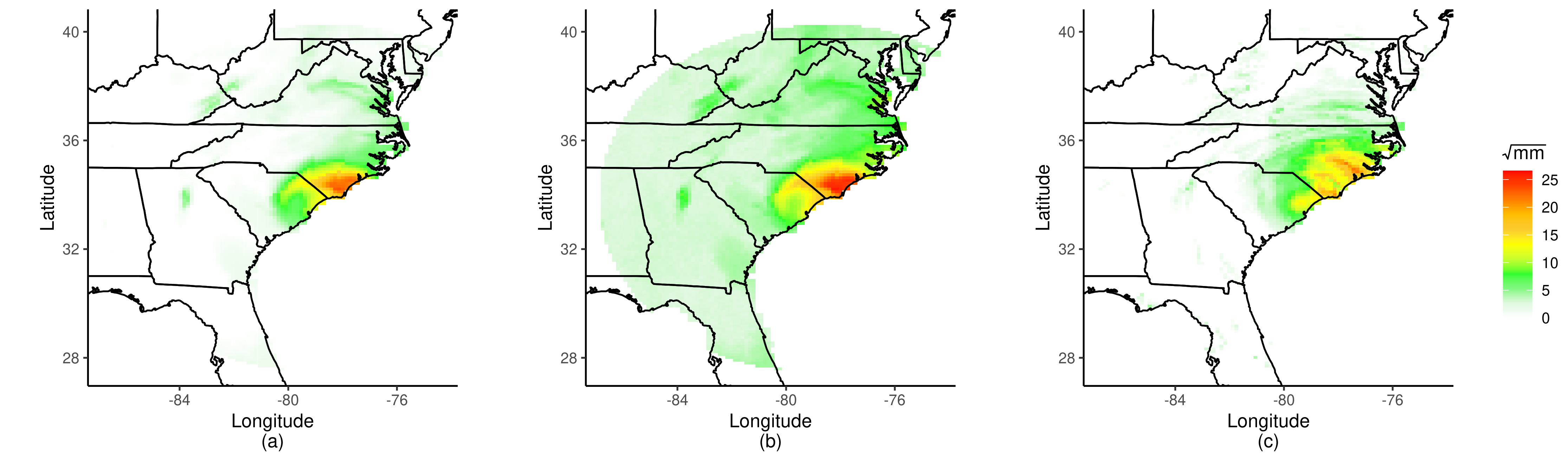}
    \caption{ Second prediction storm: Florence 2018.}
\end{figure}
\begin{figure}[ht]
    \centering
    \includegraphics[width=5.61893in]{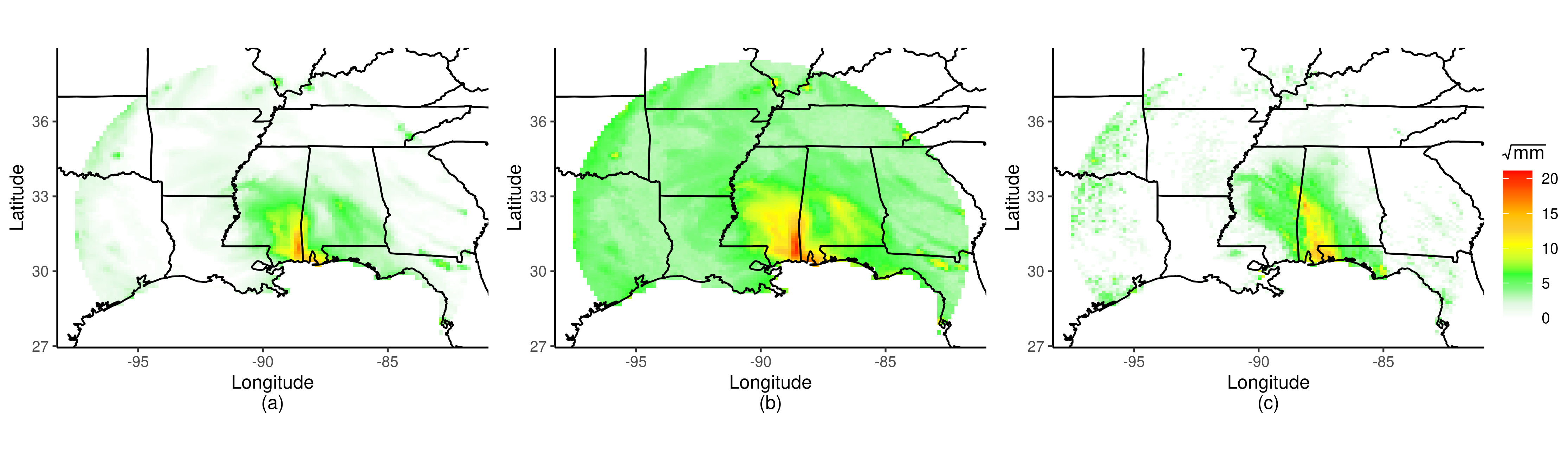}
    \caption{ Third prediction storm: Gordon 2018.}
\end{figure}
\begin{figure}[ht]
    \centering
    \includegraphics[width=5.61893in]{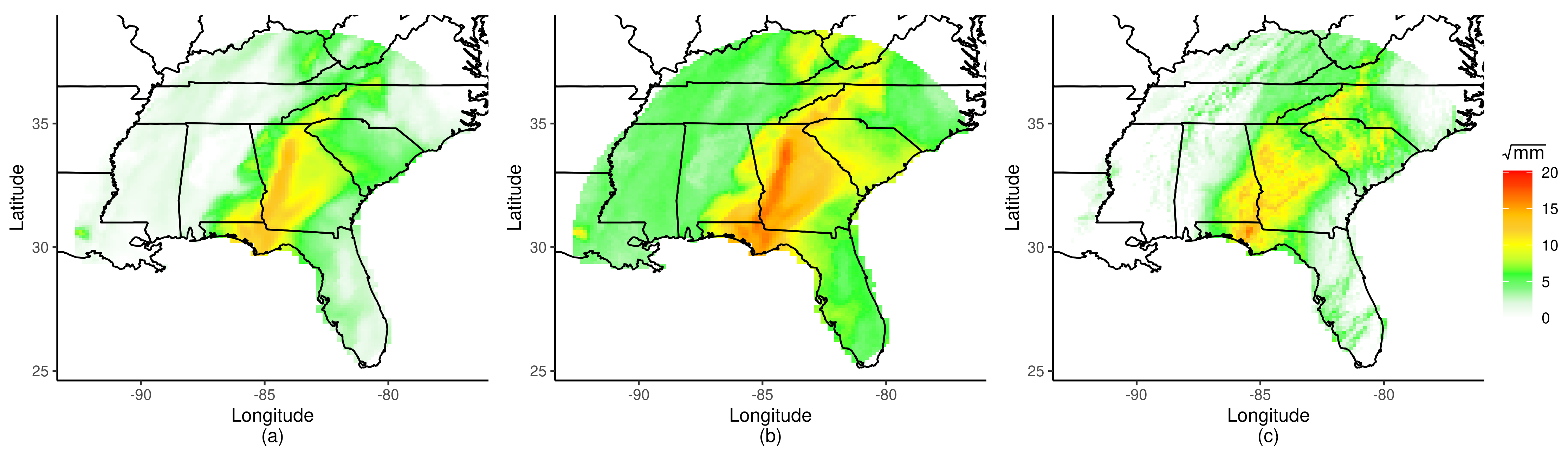}
    \caption{ Fourth prediction storm: Michael 2018.}
\end{figure}
\begin{figure}[ht]
    \centering
    \includegraphics[width=5.61893in]{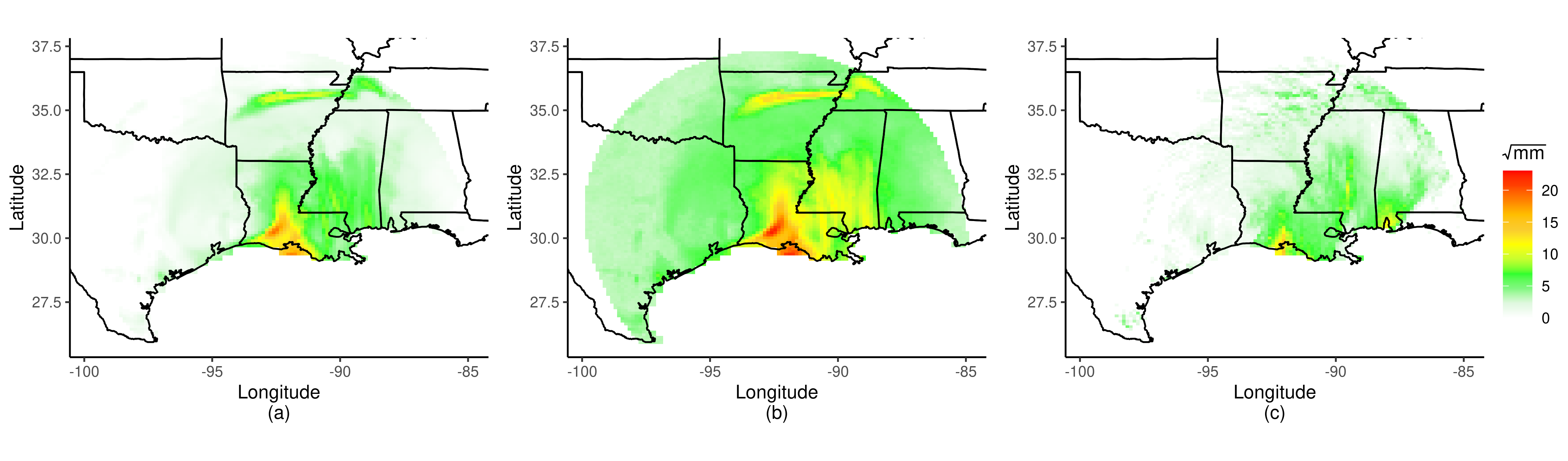}
    \caption{ Fifth prediction storm: Barry 2019.}
\end{figure}
\begin{figure}[ht]
    \centering
    \includegraphics[width=5.61893in]{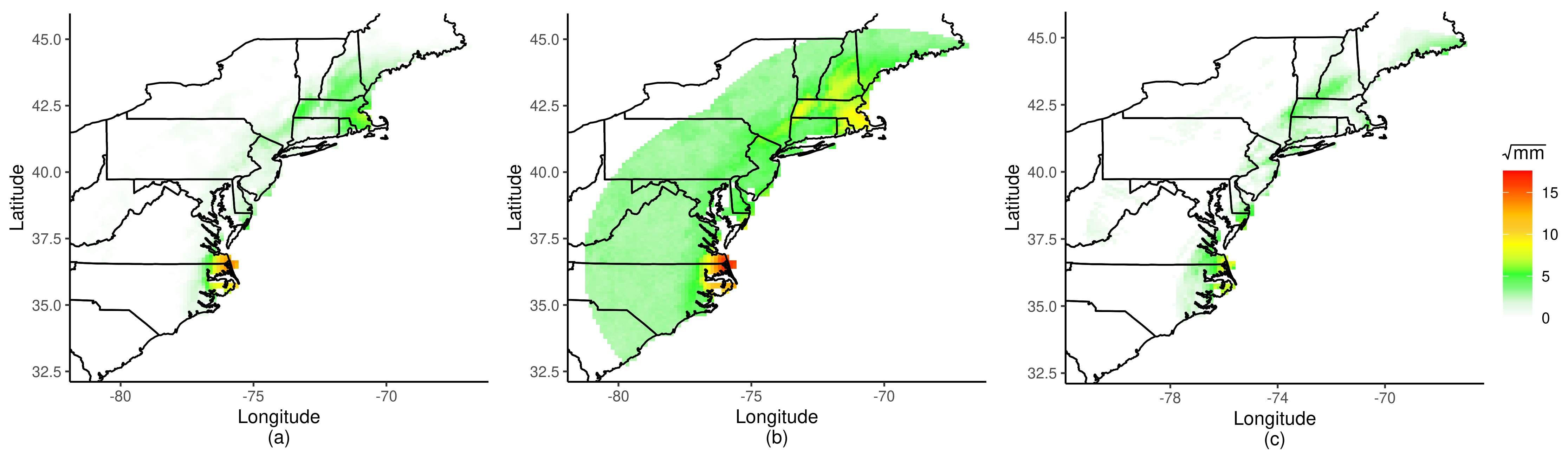}
    \caption{ Sixth prediction storm: Dorian 2019.}
\end{figure}

\begin{figure}[ht]
    \centering
    \includegraphics[width=5.61893in]{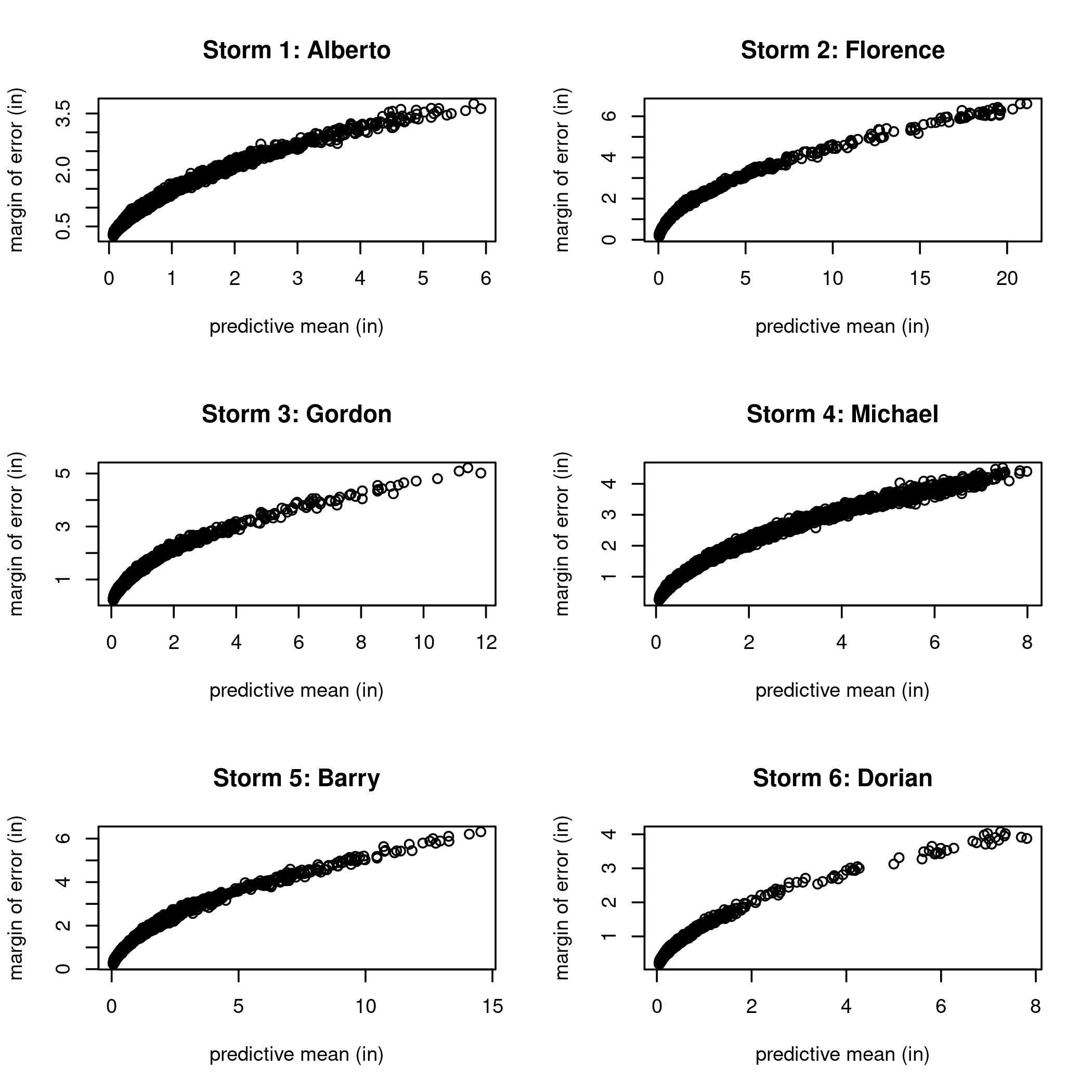}
    \caption{Margins of error for each grid point across each of the six test storms. These are calculated as half of the length of the predictive interval for each grid point of each storm.}
    \label{fig:rain_intervals}
\end{figure}

\begin{figure}
    \centering
    \includegraphics[width=5in]{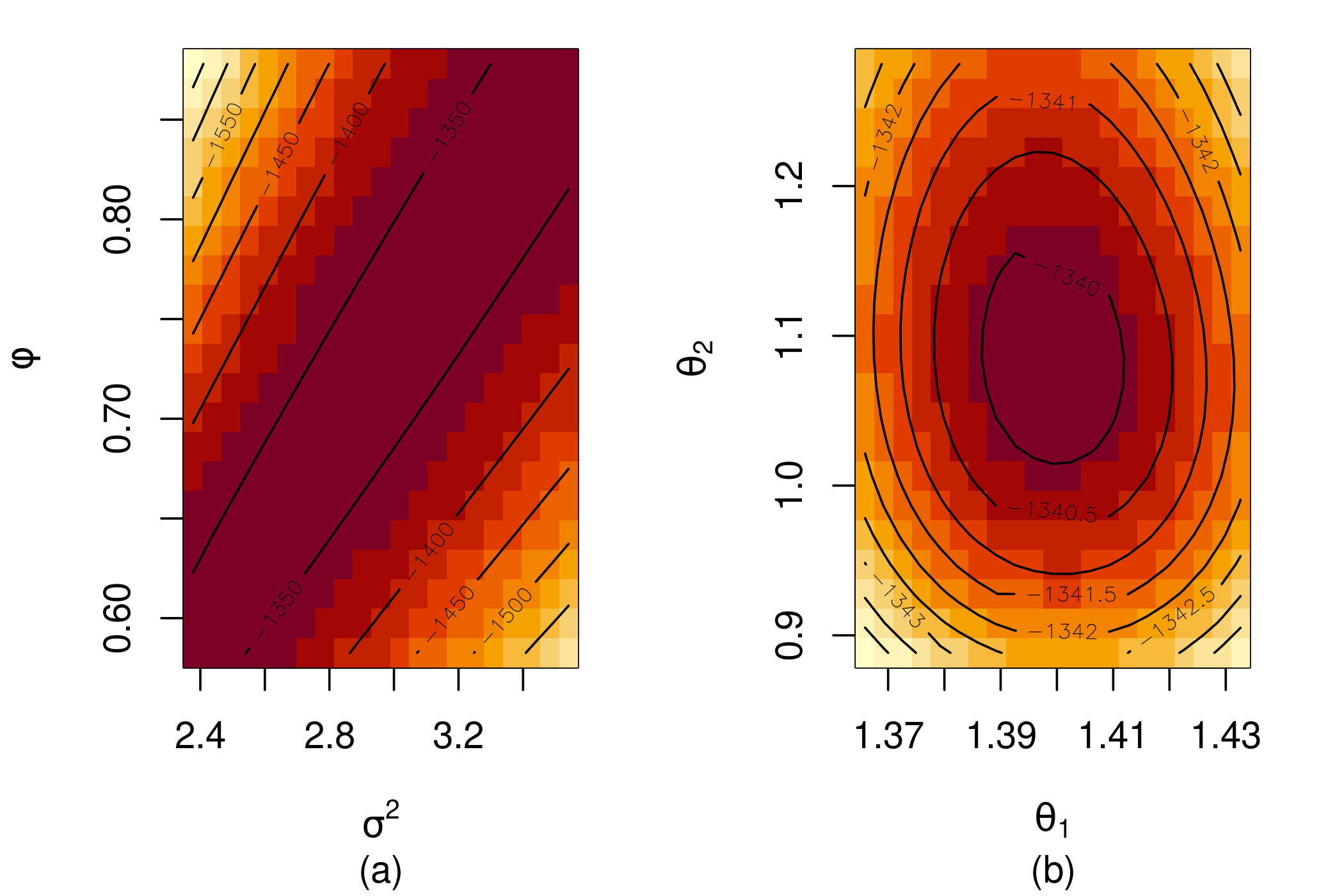}
    \caption{Log likelihood on the (a) original and (b) transformed parameter space for the first training storm. Plots are similar across all training storms.}
    \label{fig:log_likhd_grid}
\end{figure}

\clearpage 
\end{appendix}

\end{document}